%% file: main.tex
\documentclass[%
    10pt,
    aps,
    preprintnumbers,
    prd,
    noshowpacs,
    nofootinbib,
    noshowkeys,
    floatfix,
    superscriptaddress]{revtex4-2}

\usepackage[ansinew]{inputenc}
\usepackage{upgreek}
\usepackage{csquotes}
\usepackage[mathscr]{eucal}

\usepackage{amsmath}
\usepackage{amssymb}
\usepackage{mathtools}
\usepackage{cancel}
\usepackage[matha,mathx]{mathabx} 
\usepackage{isomath} 
\usepackage{physics}

\usepackage[toc,page]{appendix}
\usepackage{booktabs}
\usepackage{float}
\usepackage{wrapfig}
\usepackage{paracol} 
\usepackage{enumitem}

\usepackage{graphics,graphicx}
\usepackage{xcolor}
\usepackage[colorlinks=true,linktocpage=true,linkcolor=blue,citecolor=blue]{hyperref}
\usepackage{cleveref} 
    \crefformat{pluralequation}{eqs.~(#2#1#3)}
    \crefname{pluralequation}{Equations}{eqs.}
\usepackage{natbib}

\DeclareMathOperator{\coloneqq}{\equiv} 
\newcommand{\vecc}{\vectorarrow} 
\renewcommand{\vec}{\vectorsym} 
\newcommand{\tens}{\tensorsym} 

\DeclareMathAlphabet\mathbfcal{OMS}{cmsy}{b}{n} 

\let\oldsqrt\sqrt
\def\sqrt{\mathpalette\DHLhksqrt}
\def\DHLhksqrt#1#2{%
\setbox0=\hbox{$#1\oldsqrt{#2\,}$}\dimen0=\ht0
\advance\dimen0-0.2\ht0
\setbox2=\hbox{\vrule height\ht0 depth -\dimen0}%
{\box0\lower0.4pt\box2}}

\newcommand{\Kn}{\mathrm{Kn}}

\newcommand{\IRe}{\mathrm{IRe}} 
\newcommand{\KnIRe}{\mathrm{Kn}\, \mathrm{IRe}}

\newcommand{\fact}[1]{#1 !\,} 
\newcommand{\dfact}[1]{#1 !\mkern-1mu !\,} 

\renewcommand{\phi}{\varphi}

\begin{document}

\title{Relativistic dissipative hydrodynamics for particles of arbitrary mass}

\author{Semyon Potesnov}
\email{potesnov@itp.uni-frankfurt.de}
\affiliation{Institute for Theoretical Physics, Goethe University, Max-von-Laue-Str. 1, D-60438 Frankfurt am Main, Germany}

\author{David Wagner}
\email{david.wagner@unifi.it}
\email{dwagner@itp.uni-frankfurt.de}
\affiliation{Department of Physics \& Astronomy, University of Florence, Via G. Sansone 1, 50019 Sesto Fiorentino, Florence, Italy}
\affiliation{INFN Sezione di Firenze, Florence, Italy}

\begin{abstract}
    Employing a kinetic framework, we calculate all transport coefficients for relativistic dissipative (second-order) hydrodynamics for arbitrary particle masses in the 14-moment approximation.
    Taking the non-relativistic limit, it is shown that the relativistic theory reduces to the Grad equations computed in the \enquote{order-of-magnitude} approach.
\end{abstract}




\maketitle

\section{Introduction}\label{Introduction}

    \input{part-1}

\section{Formulation of relativistic fluid dynamics}\label{Formulation-of-relativistic-fluid-dynamics}

    \input{part-2}

\section{Bridging relativistic and non-relativistic hydrodynamics}\label{Bridging-relativistic-and-non-relativistic-hydrodynamics}

    \input{part-3}

\section{Conclusions}\label{Conclusions}

    \input{conclusions}

\section*{Acknowledgments}

    \input{acknowledgements}

\appendix
\input{appendix}

\bibliography{main}{}

\end{document}

%% file: part-1.tex
Relativistic hydrodynamics has been shown to play a central role in the description of high-energy plasmas, particularly in environments where strong energy and momentum fluxes lead to highly non-linear behavior.
Understanding the dynamics of such systems is essential for interpreting observations of relativistic jets, accretion disks, and the quark-gluon plasma produced in heavy-ion collisions.
Despite significant theoretical and numerical progress~\cite{Romatschke:2017ejr,Rezzolla:2013},
accurately modeling these plasmas remains a formidable challenge.

Theoretically, one of the most demanding aspects of relativistic plasmas is the inclusion of dissipative effects in hydrodynamic descriptions.
The incorporation of a self-consistent treatment of these dissipative processes requires going beyond the standard relativistic Euler equations to more complex formulations, such as those based on extended thermodynamics and kinetic theory~\cite{Rocha:2023}.
Even though the latter has become well-established in the high-energy community, and has given rise to many successful numerical studies of relativistic fluids, some conceptual difficulties remain, particularly in understanding the region of applicability of these frameworks, as well as ensuring causality and stability of the equations.

Experimentally, these theoretical limitations translate into practical difficulties in interpreting data from high-energy physics and astrophysical observations.
For example, in heavy-ion collisions at facilities like the Large Hadron Collider (LHC) and the Relativistic Heavy Ion Collider (RHIC), precise modeling of the evolution of the quark-gluon plasma is crucial for extracting information about transport coefficients and phase transitions~\cite{Shen:2020mgh,Heinz:2013th,Gale:2013da}.
Similarly, in astrophysical systems, where the inclusion of dissipative effects is a more recent advance~\cite{Most:2021zvc,Chabanov:2021dee,Most:2022yhe}, accurate hydrodynamic models are necessary to infer the properties of, e.g., accreting compact objects or binary neutron-star mergers.
In both cases, shortcomings in the theoretical framework limit our ability to match experimental data to the nature of underlying physical processes.

Recent developments in causal and stable formulations of dissipative hydrodynamics, coupled with advances in numerical techniques, are paving the way toward improved models of relativistic plasmas~\cite{Rocha:2023}.
By bridging the gap between kinetic theory and hydrodynamics, these approaches have the potential to offer new insights into the fundamental properties of high-energy matter.
While such hydrodynamic theories based on kinetic theory have seen tremendous progress, the relevant transport coefficients have so far been computed mostly for (almost) massless fluid constituents~\cite{Denicol:2012cn,Molnar:2013lta,Denicol:2014vaa,Rocha:2023hts,Wagner:2022ayd}.
In this paper, building upon these works, we compute all transport coefficients in the massive case, thereby presenting a kinetic framework which allows to describe relativistic dissipative hydrodynamics for arbitrary particle masses.
Furthermore, we compute the non-relativistic limit of this theory, showing that it reduces to the well-established non-relativistic hydrodynamics based on the kinetic formulation of Grad~\cite{Grad:1949zza}.

The paper is structured as follows.
\Cref{Notation-and-conventions} is dedicated to conventions and notation, and in \cref{Developing-a-transport-model-for-hydrodynamics} the essence of modeling hydrodynamics using kinetic theory is briefly summarized.
In \cref{Formulation-of-relativistic-fluid-dynamics} the framework of hydrodynamics for a (special-)relativistic monatomic fluid is established.
Specifically, \cref{Theory-of-relativistic-fluid-dynamics} showcases the general methods of fluid dynamics, and \cref{Microscopic-framework-of-kinetic-theory} introduces the essential aspects of kinetic theory.
Combining both frameworks, in \cref{Kinetic-theory-for-hydrodynamics} the evolution equations of the fluid resulting from kinetic theory are derived, and \cref{Transport-properties-of-the-fluid} showcases all transport coefficients of relativistic dissipative hydrodynamics for arbitrary mass, constituting the first major result of this work.
Then, in \cref{Bridging-relativistic-and-non-relativistic-hydrodynamics}, the second main result of this paper is presented, i.e., the establishment of the connection between the relativistic and the non-relativistic regimes of second-order hydrodynamics.
Finally, \cref{Conclusions} recapitulates our findings and concludes the paper.

\subsection{Notation and conventions}\label{Notation-and-conventions}

For convenience, we demand $c \coloneqq 1 \coloneqq k_{B}$, where $c$ is the speed of light in vacuum, and $k_B$ the Boltzmann constant.
Furthermore, we employ the metric tensor of mostly minus signature to normalize the fluid velocity $\vec{u}$ to $g_{\mu \nu} u^{\mu} u^{\nu} = 1$.
This velocity defines a distinctive observer frame, the so-called \enquote{local-rest frame} by $u^{\mu}_{\text{LR}} = (1, 0 \in \mathbb{R}^3)^T$.
Here, both the indexed and index-free notation for vectors and tensors is used, both denoted by Greek indices in Minkowski space (slanted) and by Roman in euclidean $\mathbb{R}^3$ (upright).
As such, $\sigma^{\mu \nu}$ or $\tens{\sigma}$ denotes the same $(2, 0)$-rank tensor, $V^{\mu}$ or $\vec{V}$ a 4-vector, and a 3-vector is denoted by $k_i$ or $\vecc{k}$.
Further, we denote the symmetric part of $\sigma^{\mu \nu}$ by $\sigma^{( \mu \nu )} \coloneqq \left( \sigma^{\mu \nu} + \sigma^{\nu \mu} \right) / 2$, and the antisymmetric by $\sigma^{[ \mu \nu ]} \coloneqq \left( \sigma^{\mu \nu} - \sigma^{\nu \mu} \right) / 2$.
An irreducible projection tensor $\Delta^{\mu_{1} \cdots \mu_{n}}_{\nu_{1} \cdots \nu_{\ell}}$ of rank $(n, \ell)$, satisfying
\[
    \Delta^{\mu_1 \cdots \mu_n}_{\mu_1 \cdots \mu_n} = 2n+1 ,
    \qquad
    \Delta^{\mu_i \mu_j} \Delta^{\nu_1 \cdots \nu_n}_{\mu_1 \cdots \mu_n} = 0\ \forall\ \left\{ i, j \right\} ,
    \qquad
    \Delta^{\mu_1 \cdots \mu_n}_{\nu_1 \cdots \nu_n} \Delta^{\nu_1 \cdots \nu_n}_{\lambda_1 \cdots \lambda_n} = \Delta^{\mu_1 \cdots \mu_n}_{\lambda_1 \cdots \lambda_n}
\]
is used to denote a tensor $\rho^{\langle \mu_{1} \cdots \mu_{n} \rangle} \coloneqq \Delta^{\mu_1 \cdots \mu_n}_{\nu_1 \cdots \nu_n} \rho^{\nu_{1} \cdots \nu_{n}}$, orthogonal to $u_{\mu_i} \forall i$.
One special projector of rank $(2, 0)$ onto a 3-subspace orthogonal to $\vec{u}$ is $\Delta^{\mu \nu} = g^{\mu \nu} - u^{\mu} u^{\nu}$.
Also, $\Delta^{\mu \nu}_{\alpha \beta} = \Delta^{\mu}_{( \alpha} \Delta^{\nu}_{\beta )} - \frac{1}{\Delta^{\lambda}_{\lambda}} \Delta^{\mu \nu} \Delta_{\alpha \beta}$.
Further, the basis
\[
    \left\{ u_{\mu_1}, \ldots, u_{\mu_1} \cdots u_{\mu_{n}}, \Delta^{\mu_1}_{\alpha_1}, \ldots, \Delta^{\mu_1 \cdots \mu_{n}}_{\alpha_1 \cdots \alpha_{n}} \right\}
\]
is called \enquote{irreducible} due to the projector being related to the little subgroup of the Poincar\'e group, accounting for spatial rotations of the medium.
The comoving derivative is denoted by a dot, $\dot{\tens{\pi}} \coloneqq \dd{\tens{\pi}}/\dd{\tau} \coloneqq u^\mu \partial_\mu \tens{\pi}$, and the spatial derivative reads $\nabla^\mu \coloneqq \Delta^{\mu\nu} \partial_\nu$.

The regime of large mass, where the behavior of the system is dominated by non-relativistic contributions, may be denoted by $m \rightsquigarrow \infty$.
Introducing a finite inverse temperature $\beta \coloneqq 1/T$, this is equivalent in our notation to writing $z \coloneqq \beta m$ and taking $z \rightsquigarrow \infty$.
Quantities which assume a characteristic form in this regime will be furnished with the $(\infty)$-index, e.g. the asymptotic pressure is $P_{0}^{(\infty)}$.

Finally, we denote the inverse Reynolds number $\mathrm{Re}^{-1}$ by $\IRe$.

\subsection{Developing a transport model for hydrodynamics}\label{Developing-a-transport-model-for-hydrodynamics}

The following discussion is applicable to the dynamics of a rarefied fluid, i.e., gas, which will be addressed by hydrodynamics, as usually found in literature~\cite{Landau_Lifshitz:1959}.

\paragraph{Assumptions of fluid dynamics}

Relativistic fluid dynamics, as introduced in \cref{Theory-of-relativistic-fluid-dynamics}, is an effective field theory in terms of the fluid-velocity field $\vec{u}$, which is assumed to be applicable when a separation of microscopic and macroscopic scales inside a many-body system is present.
Namely, the dynamics can be described in terms of a coarse-grained system comprised of many small, but macroscopic elements, each of which is referred to as a \enquote{fluid cell} in the following.
Thus, fluid dynamics is applicable when the microscopic scale $\lambda$ -- assumed unique for simplicity -- is much smaller than the scale of the fluid cell $L$; the region of applicability is certified by the Knudsen number $\Kn \coloneqq \lambda/L$ being small, $\Kn \ll 1$.

\paragraph{Ideal fluid and ambiguity of viscous forces}

For the so-called ideal fluid, each fluid cell is assumed to be in thermodynamic equilibrium, i.e., one has $\Kn = 0$.
However, not all fluid cells need to be in the same thermodynamic equilibrium state.
In this case, thermodynamic quantities, e.g., temperature or chemical potential, become spacetime-dependent fields, leading to the notion of local thermodynamic equilibrium (as opposed to global equilibrium, where each fluid cell is in the same maximum-entropy state)~\cite{Chiarini:2024cuv,Israel_Stewart:1979}.
The equations of motion for these fields are then provided by the conservation equations, supplemented with an equation of state.
For example, for perfect fluids the conservation equations can be derived by variational methods~\cite{Elze:1999kc,Taub:1954zz,Jackiw:2004nm}.
In contrast, the derivation of the non-equilibrium currents is still an active field of research and no variational principle has been established yet, though notable attempts~\cite{Elze:1999kc,Hongo:2024} have been made.
In general, the non-equilibrium corrections to the conserved currents in ideal fluid dynamics can be decomposed in terms of all possible projections with respect to the fluid four-velocity $\vec{u}$.
However, the resulting decomposition suffers from the intrinsic ambiguity of the relativistic fluid velocity field itself, when the existence of a preferred frame of reference, the local rest frame, inside the dissipative fluid is imposed.
One popular approach of fixing the dissipative corrections is the idea that the non-equilibrium energy flux should vanish identically in the local-rest frame.
In fluid dynamics this approach is due to \citet{Landau_Lifshitz:1959}, who introduced the fluid velocity as an eigenvector of the energy-momentum tensor.
Note that the implications of choosing different matching conditions in relativistic physics brings up discussion in the present day, in particular in the context of the so-called BNDK theory~\cite{Bemfica:2017wps,Kovtun:2019hdm}, and requires further investigation.
While, in ideal fluid dynamics, for the determination of the thermodynamic quantities, such as particle-number density, energy density, and pressure, it is sufficient to supplement the conservation equations by an equation of state, the dissipative quantities, i.e., the bulk viscous pressure $\Pi$, the particle diffusion current $\vec{V}$, and the shear-stress tensor $\vec{\pi}$, require additional information to be added.

\paragraph{Navier-Stokes and the first-order viscous corrections}

Historically, one of the first prescriptions for dissipative quantities was constructed by requiring the entropy production to be a quadratic form \cite{Landau_Lifshitz:1959,Eckart:1940te}, which results in
\begin{equation} \label{Navier-Stokes:dissipative-quantities}
    \Pi \coloneqq - \zeta \theta ,
    \qquad
    V^{\mu} \coloneqq \varkappa I^{\mu} ,
    \qquad
    \pi^{\mu \nu} \coloneqq 2 \eta \sigma^{\mu \nu} ,
\end{equation}
where the bulk viscosity $\zeta$, particle diffusion $\varkappa$ and shear viscosity $\eta$ are positive-definite coefficients.
Relations \eqref{Navier-Stokes:dissipative-quantities} are a relativistic generalization of the constitutive relations in conventional Navier-Stokes theory, and inserting them into the conservation equations of the fluid gives rise to the so-called relativistic Navier-Stokes (first-order) equations.
The resulting set of diffusion-type -- parabolic -- equations is valid up to second order in the Knudsen number, and is known to be unstable~\cite{Hiscock:1985zz}.
One way to circumvent the instability lies in promoting the dissipative quantities to dynamical variables, and deriving their evolution equations from a suitable microscopic theory to render the system hyperbolic.
These considerations lead to considering a kinetic model for the microscopic dynamics of the fluid.
    \footnote{Notably, even in the non-relativistic case for some specific scenarios the Navier-Stokes equations are known to fail, requiring a kinetic model in the region of instability~\cite[chap. 4]{Agrawal:2019}, e.g., Grad's transport equations.}

\paragraph{Microscopic interactions in the fluid}

A microscopic (kinetic)
description of the fluid (without fluctuations) will be provided in \cref{Microscopic-framework-of-kinetic-theory}.
In kinetic theory, a phase-space description of the system is provided by the single-particle distribution function $f_{\vecc{k}}$, whose evolution is determined by the Boltzmann equation.
Therein, the information about microscopic interactions is contained in the collision integral $C[f_{\vecc{k}}]$.
In particular, by virtue of the symmetries of the collision integral, which encode the conservation of particle number and energy in binary elastic collisions, one recovers the conservation equations for the particle-number current and the energy-momentum tensor~\cite{de_Groot:1980}.
Furthermore, the notion of local thermodynamic equilibrium is determined by the class of distribution functions $f_{0\vecc{k}}$ that make the collision term vanish.
Thus, only the non-equilibrium distribution function $\var{f_{\vecc{k}}} \coloneqq f_{\vecc{k}} - f_{0\vecc{k}}$ has a non-trivial collision integral.
In kinetic theory, the local equilibrium with greatest entropy, i.e., global equilibrium,
is reached by virtue of the $H$-theorem~\cite{de_Groot:1980}.
Hence, kinetic theory provides a microscopic theory including dissipation, and can thus be employed as a starting point for a viable theory of relativistic dissipative hydrodynamics.

\paragraph{Transport model for hydrodynamics}\label{intro:transport-model}

Though the regions of validity of kinetic theory and hydrodynamics have a limited overlap, in many cases a kinetic framework is sufficient to determine the transport properties of a rarefied relativistic fluid.
    \footnote{Naturally, $\Kn \mapsto 0$ has a direct interpretation of vanishing mean free path between fluid constituents.}
Indeed, the resulting fluid-dynamical theory has seen successes even when pushed to its limits in small and short-lived, as well as strongly coupled systems, such as the quark-gluon plasma formed in high-energy heavy-ion collisions~\cite{Huovinen:2001cy,Kolb:2003dz}.

\paragraph{Second-order hydrodynamics}

In order to appropriately quantify how \enquote{close} the distribution function $f_{\vecc{k}}$ must be to the local-equilibrium function $f_{0\vecc{k}}$, it is sensible to introduce the inverse Reynolds numbers through $\IRe_{i} \sim \var f_{\vecc{k}}/f_{0 \vecc{k}}$.
Explicitly, in terms of the dissipative quantities, one can define $\IRe_\Pi\coloneqq \Pi/P_0$, $\IRe_\vec{V}\coloneqq \sqrt{-V^\mu V_\mu}/(\beta P_0)$, and $\IRe_\vec{\pi}\coloneqq \sqrt{\pi^{\mu\nu}\pi_{\mu\nu}}/P_0$, with the pressure $P_0$ and the inverse temperature $\beta$.
Then, after transforming the Boltzmann equation into an infinite system of partial differential equations in terms of its moments, this tower of equations can be truncated at second order, i.e., keeping terms up to orders $\order{\Kn^2,\, \KnIRe,\, \IRe^2}$.
This truncation can be done in several ways, such as the so-called DNMR prescription~\cite{Denicol:2012cn}, or with the inverse-Reynolds-dominance (IReD) method~\cite{Wagner:2022ayd}, the latter of which will be considered in this work.
In any case, as a result one obtains a set of partial differential equations for the dissipative quantities $\Pi$, $\vec{V}$, and $\vec{\pi}$, which provide an improved description compared to the Navier-Stokes relations \eqref{Navier-Stokes:dissipative-quantities}.
Notably, IReD is distinguished by the absence of $\order{\Kn^2}$ terms, which are known to render the system of equations parabolic.
However, it retains the contributions of the absent terms by amending the values of the $\order{\KnIRe}$ terms~\cite[sec.~V]{Wagner:2022ayd}.
Due to the habit of excluding these unstable terms in numerical simulations~\cite{Denicol:2012vq} the approach of IReD should theoretically improve the accuracy of the resulting theory~\cite{Wagner:2023jgq}.

%% file: part-2.tex
In this section the relativistic theories of hydrodynamics and kinetic theory will be established, and a prescription of matching the microscopic degrees of freedom of the fluid to kinetic theory will be shown.

\subsection{Theory of relativistic fluid dynamics}\label{Theory-of-relativistic-fluid-dynamics}

Relativistic fluid dynamics is based on the conservation equations
\begin{subequations}
\begin{align}
    \label{fluid-dynamics:conservation-N}
        \partial_\mu N^\mu &= 0 \;,
    \\
    \label{fluid-dynamics:conservation-T}
        \partial_\mu T^{\mu \nu} &= 0 \;,
\end{align}
\end{subequations}
where $\vec{N}$ is the particle-number current and $\tens{T}$ the energy-momentum tensor.
\Cref{fluid-dynamics:conservation-N,fluid-dynamics:conservation-T} hold for the sum of equilibrium and non-equilibrium contributions, i.e. $\vec{N} \coloneqq \vec{N}_{0} + \var{\vec{N}}$ and $\tens{T} \coloneqq \tens{T}_{0} + \var{\tens{T}}$.
The former read
\begin{subequations}
    \label[pluralequation]{fluid-dynamics:decomposition-N-T}
\begin{align}
    \label{fluid-dynamics:equilibrium-N}
    N_{0}^{\mu}
    &= n_0 u^{\mu} \;,
    \\
    \label{fluid-dynamics:equilibrium-T}
    T_{0}^{\mu \nu}
    &= \varepsilon_0 u^{\mu} u^{\nu}
        - P_0  \Delta^{\mu \nu} \;,
\end{align}
where $n_0, \varepsilon_0$ are the thermodynamic particle- and energy densities, and $P_0$ the pressure.
Using the basis
\[
    \left\{ u^{\mu}, u^{\mu} u^{\nu}, \Delta^{\mu \nu}, \Delta^{\mu \nu}_{\alpha \beta} \right\} \;,
\]
the non-equilibrium currents are decomposed as
\begin{align}
    \label{fluid-dynamics:nonequilibrium-N}
    \var{N}^{\mu}
    &\coloneqq \var{n} u^{\mu}
        + V^{\mu} \;,
    \\
    \label{fluid-dynamics:nonequilibrium-T}
    \var{T}^{\mu \nu}
    &\coloneqq \var{\varepsilon} u^{\mu} u^{\nu}
        - \Pi \Delta^{\mu \nu}
        + 2 u^{( \mu} W^{\nu )}
        + \pi^{\mu \nu} \;,
\end{align}
\end{subequations}
where the dissipative quantities are the non-equilibrium correction to the particle-number density $\var{n}$ and the energy density $\var{\varepsilon}$, as well as the bulk viscous pressure $\Pi$, the particle-diffusion current $\vec{V}$, the energy diffusion $\vec{W}$ and the shear-stress tensor $\tens{\pi}$.

Introducing the Landau frame,
\begin{equation} \label{fluid-dynamics:Landau-frame}
    u_\mu T^{\mu \nu} \coloneqq \varepsilon_0 u^\nu \;,
    \qquad u_\mu N^\mu \coloneqq n_0 \;,
\end{equation}
imposes $\vec{W} = 0$ and $\var{\varepsilon} = 0 = \var{n}$, according to \cref{fluid-dynamics:nonequilibrium-N,fluid-dynamics:nonequilibrium-T}.
These so-called \enquote{matching conditions} are implied by choosing $\var{T_{\mu \nu}} u^{\mu} \coloneqq 0 \coloneqq \var{N_{\mu}} u^{\mu}$.

Inserting \cref{fluid-dynamics:equilibrium-N,fluid-dynamics:equilibrium-T,fluid-dynamics:nonequilibrium-N,fluid-dynamics:nonequilibrium-T} with the constraints of \cref{fluid-dynamics:Landau-frame} into \cref{fluid-dynamics:conservation-N,fluid-dynamics:conservation-T} yields the equations of motion for the fluid,
\begin{subequations}
    \label[pluralequation]{fluid-dynamics:equation-of-motions}
\begin{align}
        \label{fluid-dynamics:decomposition:eom-1}
    \partial_\mu N^\mu = 0 &\Leftrightarrow \dot{n}_0 + n_0 \theta + \partial_\mu V^\mu = 0 \;,
    \\
        \label{fluid-dynamics:decomposition:eom-2}
    u_\alpha \partial_\beta T^{\alpha \beta} = 0 &\Leftrightarrow \dot{\varepsilon}_0 + \left( \varepsilon_0 + P_0 + \Pi \right) \theta - \pi^{\alpha \beta} \sigma_{\alpha \beta} = 0 \;,
    \\
        \label{fluid-dynamics:decomposition:eom-3}
    \Delta^{\mu}_\alpha \partial_\beta T^{\alpha \beta} = 0 &\Leftrightarrow \left( \varepsilon_0 + P_0 + \Pi \right) \dot{u}^\mu - \nabla^\mu (P_0 + \Pi) + \Delta^{\mu}_{\alpha} \partial_\beta \pi^{\alpha \beta} = 0 \;,
\end{align}
\end{subequations}
where the expansion scalar $\theta \coloneqq \nabla_\mu u^\mu$, the shear tensor $\sigma^{\mu \nu} \coloneqq \nabla^{\langle \mu} u^{\nu \rangle}$, and the vorticity tensor $\omega^{\mu \nu} \coloneqq \nabla^{[ \mu} u^{\nu ]}$ were defined.
These irreducible tensors appear in the so-called relativistic Cauchy-Stokes decomposition,
\begin{equation}
        \label{Cauchy-Stokes-decomposition}
    \partial_\mu u_\nu
    = u_\mu \dot{u}_\nu
        + \frac{1}{3} \Delta_{\mu \nu} \theta
        + \sigma_{\mu \nu}
        + \omega_{\mu \nu} \;.
\end{equation}
Note that the local-equilibrium quantities are governed by the laws of thermodynamics,
\begin{subequations}
    \label{hydrodynamics:EOS}
\begin{align}
        \label{fluid-dynamics:euler-relation}
    \varepsilon_0 &= T s_0 - P_0 + \mu n_0 \;, \\
    \dd{s_0} &= \beta \dd{\varepsilon_0} - \alpha \dd{n_0} \;, \\
        \label{fluid-dynamics:duhem-gibbs-relation}
    \dd{P_0} &= s_0 \dd{T} + n_0 \dd{\mu} \;,
\end{align}
where $\beta$ denotes the inverse temperature, $\alpha \coloneqq \beta \mu$ the thermal potential, and the entropy density $s_0$ has been introduced.
Hence, the intensive quantities can be written as
\begin{equation}
        \label{fluid-dynamics:maxwell-relations}
    \beta = \pdv{s_0}{{\varepsilon_0}} \eval_{n_0} , \quad
        \alpha = - \pdv{s_0}{{n_0}} \eval_{\varepsilon_0} , \quad
        s_0 = \pdv{P_0}{T} \eval_{\mu} , \quad
        n_0 = \pdv{P_0}{\mu} \eval_{T} \;.
\end{equation}
\end{subequations}

The fluid-dynamical \cref{fluid-dynamics:decomposition:eom-1,fluid-dynamics:decomposition:eom-2,fluid-dynamics:decomposition:eom-3} constitute a set of 5 partial differential equations (supplemented by an equation of state), with the solution set consisting of 14 unknown fields.
Thus, dissipative fluid dynamics is severely underdetermined and requires additional information for the determination of the 9 unknowns of $\left\{ \Pi, \vec{V}, \tens{\pi} \right\}$.
By employing kinetic theory as a microscopic formulation this set will be closed.

\subsection{Microscopic framework of kinetic theory}\label{Microscopic-framework-of-kinetic-theory}

The relativistic Boltzmann equation,
\begin{equation} \label{boltzmann:RBE}
    k^\mu \partial_\mu f_{\vecc{k}} = C[f_{\vecc{k}}] \;,
\end{equation}
is a non-linear partial integro-differential equation for the (one-particle) distribution function $f_{\vecc{k}} = f_{\vecc{k}}(x)$, where $k^\mu$ is the on-shell four-momentum of a particle $k^\mu \coloneqq (k_0 , \vecc{k} \in \mathbb{R}^3)^T$, $k_{0} \coloneqq \sqrt{ \vecc{k}^2 + m^2}$, and $\partial_\mu \coloneqq (\partial_t , \nabla)$;
the right-hand side of \cref{boltzmann:RBE} is determined by the collision integral $C[f_{\vecc{k}}]$, which we take to include binary elastic collisions only, i.e.,
\begin{equation}
    C[f_{\vecc{k}}] \coloneqq \frac{1}{2} \int \dd{K_1} \dd{K_2} \dd{K'} W_{\vecc{k}\vecc{k}'\to \vecc{k}_1\vecc{k}_2} \left(f_{\vecc{k}_1} f_{\vecc{k}_2} \widetilde{f}_{\vecc{k}}\widetilde{f}_{\vecc{k}'}-f_{\vecc{k}'}f_{\vecc{k}}\widetilde{f}_{\vecc{k}_1}\widetilde{f}_{\vecc{k}_2}\right)\;,\label{eq:def_C}
\end{equation}
where $\dd{K} \coloneqq \dd[3]{\vecc{k}} / [(2\pi)^3 k^0]$ is the Lorentz-invariant momentum-space measure, and
\begin{equation}
    W_{\vecc{k}\vecc{k}'\to \vecc{k}_1\vecc{k}_2} \coloneqq  s\sigma (2\pi)^5 \delta^{(4)}(k+k'-k_1-k_2)
\end{equation}
is the transition rate, with the Mandelstam variable $s\coloneqq(k+k')^2$ and the total cross section $\sigma$, which we take to be constant.
The function $\widetilde{f}_{0\vecc{k}} \coloneqq 1 - a f_{0\vecc{k}}$ is introduced to incorporate the effect of Bose-enhancement for $a = -1$, whereas the effect of Pauli-blocking is taken into account for $a = +1$.
For classical particles, one has $a \to 0$.

The equilibrium distribution function is given by~\cite{de_Groot:1980}
\begin{equation}
        \label{boltzmann:distribution-function}
    f_{0\vecc{k}} \coloneqq \left[ \exp \left( \beta E_{\vecc{k}} - \alpha \right) + a \right]^{-1} \;,
\end{equation}
Notice that $E_{\vecc{k}} \coloneqq u^{\mu} k_{\mu}$ is determined by the LR frame defined through the four-velocity.
Using this definition, the orthogonal projection of the four-momentum takes the form $k^{\langle \mu \rangle} = k^{\mu} - E_{\vecc{k}} u^{\mu}$.
As we will see further on, since the function $f_{0\vecc{k}}$ contains the (microscopic) equilibrium information, all thermodynamic quantities can be expressed via the so-called thermodynamic integrals,
\begin{subequations}
    \label{thermodynamic:IJ}
\begin{align}
        \label{thermodynamic:I}
    I_{nq}
    &\coloneqq \frac{(-1)^q}{(2q+1)!!} \left\langle E^{n-2q}_{\vecc{k}} \left( \Delta^{\alpha \beta} k_\alpha k_\beta \right)^q \right\rangle_{0}
    \coloneqq \frac{(-1)^q}{(2q+1)!!} \int \dd{K}\, f_{0\vecc{k}} E^{n-2q}_{\vecc{k}} \left( \Delta^{\alpha \beta} k_\alpha k_\beta \right)^q \;, \\
        \label{thermodynamic:J}
    J_{nq}
    &\coloneqq \pdv{I_{nq}}{{\alpha}} \eval_{\beta} \quad \forall\ q > - 3 / 2 \;,
\end{align}
\end{subequations}
Note that for a classical fluid it holds that $I_{nq} = J_{nq}$.
The following thermodynamic functions will also be of use:
\begin{subequations}
    \label{thermodynamic:GD}
\begin{align}
    G_{nm} &\coloneqq J_{n0} J_{m0} - J_{n-1,0} J_{m+1,0} \;, \\
    D_{nq} &\coloneqq J_{n+1,q} J_{n-1,q} - \left( J_{nq} \right)^2 \;.
\end{align}
\end{subequations}
The deviation of the distribution function from equilibrium is defined as $\delta f_{\vecc{k}} \coloneqq f_{\vecc{k}} - f_{0\vecc{k}}$.
By introducing the irreducible moments of $\delta f_{\vecc{k}}$,
\begin{equation}
        \label{boltzmann:irreducible-moments}
    \rho^{\mu_1 \cdots \mu_\ell}_{n}
    \coloneqq \left\langle E^{n}_{\vecc{k}} k^{\langle \mu_1} \cdots k^{\mu_\ell \rangle} \right\rangle_\delta
    \coloneqq \int \dd{K} \delta f_{\vecc{k}} E^{n}_{\vecc{k}} k^{\langle \mu_1} \cdots k^{\mu_\ell \rangle} \;,
\end{equation}
one can write the distribution function as~\cite{Denicol:2012cn}
\begin{equation}
    f_{\vecc{k}}
    = f_{0 \vecc{k}} \left( 1 + \widetilde{f}_{0\vecc{k}} \sum^{\infty}_{\ell=0} \sum^{N_\ell}_{n=0} \mathcal{H}^{(\ell)}_{\vecc{k} n} \rho^{\mu_1 \cdots \mu_\ell}_{n} k_{\langle \mu_1} \cdots k_{\mu_\ell \rangle} \right) \;.
\end{equation}
Here, the thermodynamic functions $\mathcal{H}^{(\ell)}_{\vecc{k} n}$ are given by a series of polynomials in terms of $E_{\vecc{k}}$.
    \footnote{In the ultra-relativistic case, the polynomials $\mathcal{H}_{{\vecc{k}}n}^{(\ell)}$ are given by a sum of associated Laguerre polynomials.
    It should be noted that, at least in this case, the moment expansion presupposes that the function $\phi_{\vecc{k}} \coloneqq f_{\vecc{k}} / f_{0{\vecc{k}}}$ is square-integrable, which is not always true, as shown recently in Ref.~\cite{deBrito:2024qow}.}
The numbers $N_\ell$ characterize the truncation order of the moment expansion, with the full distribution function being recovered when $N_\ell\to \infty \,\forall \,\ell$.
In this work, for simplicity we will employ the 14-moment expansion, parametrized by $\left\{ N_{0} = 2, N_{1} = 1, N_{2} = 0 \right\}$.
Note that the irreducible moments of negative order can be represented as
\begin{equation}
    \rho_{-r}^{\mu_1\cdots \mu_\ell} =\sum_{n=0}^{N_\ell}\mathcal{F}^{(\ell)}_{rn} \rho_n^{\mu_1\cdots \mu_\ell}\;,
\end{equation}
with the coefficients
\begin{equation}
    \mathcal{F}^{(\ell)}_{rn}\coloneqq \frac{(-1)^\ell\ell!}{(2\ell+1)!!} \int \dd{K} f_{0\vecc{k}}\widetilde{f}_{0\vecc{k}} E_{\vecc{k}}^{-r}\mathcal{H}^{(\ell)}_{\vecc{k}n} \left(E_{\vecc{k}}^2-m^2\right)\;.\label{eq:rho_r<0}
\end{equation}

Similarly, one introduces the irreducible moments of the collision term,
\begin{equation}
        \label{boltzmann:collision-term}
    C^{\langle \mu_1 \cdots \mu_\ell \rangle}_{n}
    \coloneqq \int \dd{K} E_{\vecc{k}}^{n} k^{\langle \mu_1} \cdots k^{\mu_\ell \rangle} C[f_{\vecc{k}}] \;,
\end{equation}
which will appear in the evolution equations for the moments $\rho_r^{\mu_1\cdots\mu_\ell}$.

\subsection{Kinetic theory for hydrodynamics}\label{Kinetic-theory-for-hydrodynamics}

\subsubsection{Matching fluid-dynamical variables to kinetic description}\label{Matching-fluid-dynamical-variables-to-kinetic-description}

The conserved currents can be written as moments of the distribution function:
\begin{subequations}
\begin{align}
        \label{kinetic-theory:conservation-N}
    N^\mu &\coloneqq \langle k^\mu \rangle \equiv \langle k^\mu\rangle_0 + \langle k^\mu \rangle_\delta \; ,
    \\
        \label{kinetic-theory:conservation-T}
    T^{\mu \nu} &\coloneqq \langle k^\mu k^\nu \rangle \equiv \langle k^\mu k^\nu \rangle_0 + \langle k^\mu k^\nu \rangle_\delta \;.
\end{align}
\end{subequations}
Due to the form of the conserved currents of fluid dynamics in \cref{fluid-dynamics:equilibrium-N,fluid-dynamics:equilibrium-T,fluid-dynamics:nonequilibrium-N,fluid-dynamics:nonequilibrium-T}, it is possible to identify the fluid-dynamical variables by
\begin{equation}
    n_0 + \var{n} = \left\langle E_{\vecc{k}} \right\rangle \;,
    \quad \varepsilon_0 + \var{\varepsilon} = \left\langle E^{2}_{\vecc{k}} \right\rangle \;,
    \quad P_0 + \Pi = - \frac{1}{3} \left\langle \Delta^{\alpha \beta} k_\alpha k_\beta \right\rangle \;,
    \quad V^\mu = \left\langle k^{\langle \mu \rangle} \right\rangle \;,
    \quad \pi^{\mu \nu} = \left\langle k^{\langle \mu} k^{\nu \rangle} \right\rangle \;.
\end{equation}
At this point, the variables are not yet related to the equilibrium frame.
Employing \cref{thermodynamic:I,boltzmann:irreducible-moments,fluid-dynamics:Landau-frame}, one can identify the local-equilibrium quantities $n_0 = I_{10}$, $\varepsilon_0 = I_{20}$, and $P_0 = I_{21}$,
    \footnote{Notice that the kinetic representation of the pressure and the particle-number density is consistent with \cref{fluid-dynamics:maxwell-relations}, $n_0 = T^{-1} \partial P_0/\partial \alpha \big|_T = \beta P_0 $, i.e., the equation of state for an ideal gas.}
as well as the dissipative ones
\begin{equation}
    \var{n} = \rho_1 = 0 \;,
    \quad \var{\varepsilon} = \rho_2 = 0 \;,
    \quad W^{\mu} = \rho_{1}^{\mu} = 0 \;,
    \quad \Pi = - \frac{m^2}{3} \rho_0 \;,
    \quad V^{\mu} = \rho_{0}^{\mu} \;,
    \quad \pi^{\mu \nu} = \rho_{0}^{\mu \nu} \;.
\end{equation}

\subsubsection{Equations of irreducible moments}\label{Equations-of-irreducible-moments}

By decomposing the Boltzmann \cref{boltzmann:RBE} in terms of the irreducible basis,
\begin{equation}
        \label{boltzmann:dotdeltaf}
    \var{\dot{f_{\vecc{k}}}}
    = E^{-1}_{\vecc{k}} C[f] - \dot{f}_{0\vecc{k}} - E^{-1}_{\vecc{k}} k^{\langle \mu \rangle} \nabla_\mu ( f_{0\vecc{k}} + \var{f_{\vecc{k}}} ) \;,
\end{equation}
and substituting the decomposition into the (projected) comoving derivative of \cref{boltzmann:irreducible-moments}, i.e.,
\begin{equation*}
        \label{boltzmann:irreducible-moments-eq}
    \dot{\rho}^{\langle \mu_1 \cdots \mu_\ell \rangle}_r
    = \Delta^{\mu_1 \cdots \mu_\ell}_{\alpha_1 \cdots \alpha_\ell} \int \dd{K}  \dv{}{{\tau}} \left( E^{r}_{\vecc{k}}\, k^{\langle \alpha_1} \cdots k^{\alpha_\ell \rangle}\, \var{f_{\vecc{k}}} \right) \;,
\end{equation*}
the latter equation may be written as
\begin{multline} \label{general-moment-equation}
    \dot{\rho}^{\langle \mu_1 \cdots \mu_\ell \rangle}_{r}
        - C_{r-1}^{\langle \mu_1 \cdots \mu_\ell \rangle}
    = \Delta^{\mu_1 \cdots \mu_\ell}_{\alpha_1 \cdots \alpha_\ell} \left\langle k^{\langle \alpha_1} \cdots k^{\alpha_\ell \rangle} \dv{}{{\tau}} E_{\vecc{k}}^{r} \right\rangle_\delta
        + \Delta^{\mu_1 \cdots \mu_\ell}_{\alpha_1 \cdots \alpha_\ell} \left\langle E_{\vecc{k}}^{r} \dv{}{{\tau}} k^{\langle \alpha_1} \cdots k^{\alpha_\ell \rangle} \right\rangle_\delta \\
        - \Delta^{\mu_1 \cdots \mu_\ell}_{\alpha_1 \cdots \alpha_\ell} \int \dd{K} E_{\vecc{k}}^{r} k^{\langle \alpha_1} \cdots k^{\alpha_\ell \rangle} \left[ \dot{f}_{0\vecc{k}} + E^{-1}_{\vecc{k}} k^{\langle \mu \rangle} \nabla_\mu ( f_{0\vecc{k}} + \var{f_{\vecc{k}}} ) \right] \;,
\end{multline}
where the irreducible moments of the collision term $C_{r-1}^{\langle \mu_1 \cdots \mu_\ell \rangle}$ were defined in \cref{boltzmann:collision-term}.
Here, the comoving derivative of $f_{0\vecc{k}} = f_{0\vecc{k}} \left( \alpha, \beta, E_{\vecc{k}} \right)$ is determined by the conservation \cref{fluid-dynamics:conservation-N,fluid-dynamics:conservation-T}, which can be reformulated in terms of $(\alpha, \beta)$ as
\begin{subequations}
    \label{fluid-dynamics:evolution-primary-fluid-dynamical-variables}
\begin{align}
        \label{alpha-eom}
    \dot{\alpha}
    &= \frac{1}{D_{20}} \left\{ -J_{30} ( n_0 \theta + \partial_\mu V^\mu ) + J_{20} \left[ ( \varepsilon_0 + P_0 + \Pi ) \theta - \pi^{\alpha \beta} \sigma_{\alpha \beta} \right] \right\} \;, \\
        \label{beta-eom}
    \dot{\beta}
    &= \frac{1}{D_{20}} \left\{ -J_{20} ( n_0 \theta + \partial_\mu V^\mu ) + J_{10} \left[ ( \varepsilon_0 + P_0 + \Pi ) \theta - \pi^{\alpha \beta} \sigma_{\alpha \beta} \right] \right\} \;, \\
        \label{u-eom}
    \dot{u}^\mu
    &= \frac{ 1 }{ \varepsilon_0 + P_0 } \left( F^\mu - \Pi \dot{u}^\mu + \nabla^\mu \Pi - \Delta^{\mu}_{\alpha} \partial_\beta \pi^{\alpha \beta} \right) \;,
\end{align}
\end{subequations}
where the pressure gradient $F^\mu\coloneqq \nabla^\mu P_0$ was introduced~\cite[Chap.~5]{Denicol_Rischke:2021}.
As shown in Ref.~\cite{deBrito:2024vhm}, a general equation of motion for the irreducible moment of arbitrary tensor rank $\ell \geq 0$ can be derived from \cref{general-moment-equation}; here only the first three moment equations~\cite{Denicol:2012cn} are presented,
\begin{subequations}
    \label{kinetic:equations-of-irreducible-moments}
\begin{align}
    \label{eq:rho}
    \begin{split}
    \dot{\rho}_r - C_{r-1}
    ={}& \alpha^{(0)}_r \theta
        - \frac{G_{2r}}{D_{20}} \Pi \theta
        + \frac{G_{2r}}{D_{20}} \pi^{\mu \nu} \sigma_{\mu \nu}
        + \frac{G_{3r}}{D_{20}} \partial_\mu V^\mu
        + (r-1) \rho^{\mu \nu}_{r-2} \sigma_{\mu \nu} + r \rho^{\mu}_{r-1} \dot{u}_\mu \\
        & - \nabla_\mu \rho^{\mu}_{r-1}
        - \frac{1}{3} \big[ (r+2) \rho_{r} - (r-1) m^2 \rho_{r-2} \big] \theta \;,
    \end{split} \\
    \label{eq:rho_mu}
    \begin{split}
    \dot{\rho}^{\langle \mu \rangle}_r - C^{\langle \mu \rangle}_{r-1}
    ={}& \alpha^{(1)}_{r} I^\mu
        + \rho^{\nu}_{r} \omega^{\mu}{}_{\nu}
        + \frac{1}{3} \big[ (r-1) m^2 \rho^{\mu}_{r-2} - (r+3) \rho^{\mu}_{r} \big] \theta
        - \Delta^{\mu}_{\lambda} \nabla_\nu \rho^{\lambda \nu}_{r-1}
        + r \rho^{\mu\nu}_{r-1} \dot{u}_\nu \\
        & + \frac{1}{5} \big[ (2r-2) m^2 \rho^{\nu}_{r-2} - (2r+3) \rho^{\nu}_{r} \big] \sigma^{\mu}{}_{\nu}
        + \frac{1}{3} \big[ m^2 r \rho_{r-1} - (r+3) \rho_{r+1} \big] \dot{u}^\mu \\
        & + \frac{\beta J_{r+2,1}}{\varepsilon_0 + P_0} \left(
            \Pi \dot{u}^\mu
            - \nabla^\mu \Pi
            + \Delta^{\mu}_{\nu} \partial_\lambda \pi^{\lambda\nu} \right)
        - \frac{1}{3} \nabla^\mu \big( m^2 \rho_{r-1} - \rho_{r+1} \big)
        + (r-1) \rho^{\mu \nu\lambda}_{r-2} \sigma_{\lambda\nu} \;,
    \end{split} \\
    \label{eq:rho_munu}
    \begin{split}
    \dot{\rho}^{\langle \mu \nu \rangle}_{r}
        - C^{\langle \mu \nu \rangle}_{r-1}
    ={}& 2 \alpha^{(2)}_{r} \sigma^{\mu \nu}
        - \frac{2}{7} \left[
            (2r+5) \rho^{\lambda \langle \mu}_{r}
            - 2 m^2 (r-1) \rho^{\lambda \langle \mu}_{r-2} \right] \sigma^{\nu \rangle}{}_{\lambda}
        + 2 \rho^{\lambda \langle \mu}_{r} \omega^{\nu \rangle}{}_{\lambda} \\
       & + \frac{2}{15} \left[
            (r+4) \rho_{r+2}
            - (2r+3) m^2 \rho_{r}
            + (r-1) m^4 \rho_{r-2}
            \right] \sigma^{\mu \nu}
        + \frac{2}{5} \nabla^{\langle \mu} \left(
            \rho^{\nu \rangle}_{r+1}
            - m^2 \rho^{\nu \rangle}_{r-1} \right) \\
        & - \frac{2}{5} \left[
            (r+5) \rho^{\langle \mu}_{r+1}
            - m^2 r \rho^{\langle \mu}_{r-1}
             \right] \dot{u}^{\nu \rangle}
        - \frac{1}{3} \left[
            (r+4) \rho^{\mu \nu}_{r}
            - m^2 (r-1) \rho^{\mu \nu}_{r-2} \right] \theta \\
        & + (r-1) \rho^{\mu \nu \lambda \alpha}_{r-2} \sigma_{\lambda \alpha}
        - \Delta^{\mu \nu}_{\alpha \beta} \nabla_\lambda \rho^{\lambda \alpha \beta}_{r-1}
        + r \rho^{\lambda \mu \nu}_{r-1} \dot{u}_\lambda \;,
\end{split}
\end{align}
\end{subequations}
where we introduced $I^\mu \coloneqq \nabla^\mu \alpha_0$ and the first-order coefficients
\begin{subequations}
    \label{moments:thermodynamic-coefficients}
\begin{align}
    \alpha^{(0)}_r &\coloneqq (1-r) I_{r1} - I_{r0} - \frac{1}{D_{20}} \left[ G_{2r} (\varepsilon_0 + P_0 ) - G_{3r} n_0 \right] \;, \\
    \alpha^{(1)}_r &\coloneqq J_{r+1,1} - h^{-1}_{0} J_{r+2,1} \;, \\
    \alpha^{(2)}_{r} &\coloneqq I_{r+2,1} + (r-1) I_{r+2,2} \;,
\end{align}
\end{subequations}
as well as the enthalpy per particle $h_0 \coloneqq (\varepsilon_0+P_0)/n_0$.
The moments of the collision term can be separated into linear and quadratic parts~\cite{Denicol_Rischke:2021,Molnar:2013lta},
\begin{subequations}
\begin{align}
    C_{r-1} &= -\sum_{n=0}^{N_0} \mathcal{A}^{(0)}_{rn} \rho_n + N_{r-1} \;, \label{eq:coll_0}\\
    C_{r-1}^{\langle\mu\rangle} &= -\sum_{n=0}^{N_1} \mathcal{A}^{(1)}_{rn} \rho_n^\mu + N_{r-1}^{\langle\mu\rangle} \;, \label{eq:coll_1}\\
    C_{r-1}^{\langle\mu\nu\rangle} &= -\sum_{n=0}^{N_2} \mathcal{A}^{(2)}_{rn} \rho_n^{\mu\nu} + N_{r-1}^{\langle\mu\nu\rangle} \;, \label{eq:coll_2}
\end{align}
\end{subequations}
which are defined explicitly in \cref{app:coll}.

\subsubsection{Closing the system of equations}\label{Inverse-Reynolds-Dominance}

The moment \cref{eq:rho,eq:rho_mu,eq:rho_munu} can be used to obtain equations of motion for the dissipative degrees of freedom $\Pi$, $\vec{V}$, and $\tens{\pi}$.
To achieve this, the so-called inverse-Reynolds-dominance method~\cite{Wagner:2022ayd},
    \footnote{One can also get expressions for the transport coefficients via the so-called DNMR formalism of Ref.~\cite{Denicol:2012cn}, which relies on different asymptotic relations for the irreducible moments.
    The resulting transport coefficients agree in the 14-moment approximation which we use in this paper, but differ at higher orders~\cite{Wagner:2023jgq, Ambrus:2024qsa, Wagner:2023joq}.}
which is a relativistic version of the order-of-magnitude approach shown in Ref.~\cite{Struchtrup:2004},
will be used to reduce the infinite-dimensional system of moment equations.
This framework relies on a power-counting scheme in terms of the Knudsen number $\Kn \coloneqq \lambda_{\text{mfp}}/L_{\text{hydro}}$, with the mean free path $\lambda_{\text{mfp}}\coloneqq 1/(n_0 \sigma)$, and the inverse Reynolds numbers
\begin{equation} \label{IRe_Pi-V-pi}
    \IRe_{\Pi} \coloneqq \frac{\Pi}{P_0} \;,
    \quad
    \IRe_{\vec{V}} \coloneqq \frac{\sqrt{- V^{\mu} V_{\mu}}}{\beta P_0} \;,
    \quad
    \IRe_{\tens{\pi}} \coloneqq \frac{\sqrt{\pi^{\mu \nu} \pi_{\mu \nu}}}{P_0} \;,
\end{equation}
which are assumed to be small.
Assuming the Knudsen and inverse Reynolds numbers to be of the same order of magnitude to approximate the irreducible moments by the Navier-Stokes solutions, one can find the following asymptotic relations:
\begin{equation}
        \label{eq:rels}
    \rho_r
    \simeq -\frac{3}{m^2}\frac{\zeta_r}{\zeta} \Pi \;,
    \quad
    \rho_r^\mu
    \simeq \frac{\varkappa_r}{\varkappa} V^\mu \;,
    \quad
    \rho_r^{\mu\nu}
    \simeq \frac{\eta_r}{\eta}\pi^{\mu\nu} \;.
\end{equation}
Here, the symbol \enquote{$\simeq$} denotes equivalence up to first order in Knudsen and inverse Reynolds numbers, i.e., excluding terms of order $\order{\Kn^2,\KnIRe,\IRe^2}$.
Moreover, we defined
\begin{equation}
    \zeta_r \coloneqq \frac{m^2}{3}\sum_{n=0,\neq 1,2}^{N_0} \tau_{rn}^{(0)} \alpha_n^{(0)} \;,
    \quad
    \varkappa_r \coloneqq \sum_{n=0,\neq 1}^{N_1} \tau_{rn}^{(1)} \alpha_n^{(1)} \;,
    \quad
    \eta_r \coloneqq \sum_{n=0}^{N_2} \tau_{rn}^{(2)} \alpha_n^{(2)} \;,
\end{equation}
with $\tau^{(\ell)}$ being the inverse of the linearized collision matrix $\mathcal{A}^{(\ell)}$, and $\zeta \coloneqq \zeta_0$, $\varkappa \coloneqq \varkappa_0$, $\eta \coloneqq \eta_0$ the transport coefficients of the Navier-Stokes relations \eqref{Navier-Stokes:dissipative-quantities}.
Combining \cref{eq:rels,eq:rho_r<0}, the moments of negative order can be approximated by
\begin{equation}
    \rho_{-r}\simeq -\frac{3}{m^2} \gamma^{(0)}_r \Pi \;,\quad \rho_{-r}^\mu \simeq \gamma^{(1)}_r V^\mu \;,\quad \rho_{-r}^{\mu\nu} \simeq \gamma^{(2)}_r \pi^{\mu\nu}\;,
\end{equation}
where we introduced
\begin{equation}
    \gamma^{(0)}_r
    \coloneqq \sum_{n=0,\neq 1,2}^{N_0} \mathcal{F}^{(0)}_{rn} \frac{\zeta_n}{\zeta} \;,\quad \gamma^{(1)}_r
    \coloneqq \sum_{n=0,\neq 1}^{N_1} \mathcal{F}^{(1)}_{rn} \frac{\varkappa_n}{\varkappa} \;,\quad \gamma^{(2)}_r
    \coloneqq \sum_{n=0}^{N_2} \mathcal{F}^{(2)}_{rn} \frac{\eta_n}{\eta}\;.\label{eq:def_gamma}
\end{equation}

Since the relations \eqref{eq:rels} are accurate to first order in Knudsen and inverse Reynolds numbers, they can subsequently be inserted into the second-order terms of \cref{eq:rho,eq:rho_mu,eq:rho_munu}, while terms of third and higher orders are neglected.
In this way, one can close the system of moment equations, and we arrive at the following relaxation-type equations:
\begin{subequations}
    \label{eqs:transport_eqs}
\begin{align}
        \label{transport:bulk:rel}
    \tau_\Pi \dot{\Pi} + \Pi
    &= -\zeta \theta + \mathcal{J} + \mathcal{R} \;, \\
        \label{transport:diffusion:rel}
    \tau_{V} \dot{V}^{\langle\mu\rangle} + V^\mu
    &= \varkappa I^\mu + \mathcal{J}^{\mu} + \mathcal{R}^\mu \;, \\
        \label{transport:shear:rel}
    \tau_\pi \dot{\pi}^{\langle\mu \nu\rangle} + \pi^{\mu \nu}
    &= 2 \eta \sigma^{\mu \nu} + \mathcal{J}^{\mu \nu} + \mathcal{R}^{\mu \nu} \;.
\end{align}
\end{subequations}
Here, the quantities $\mathcal{J}$, $\mathcal{J}^\mu$, and $\mathcal{J}^{\mu\nu}$ contain terms of orders $\order{\KnIRe_{i}}$, while the quantities $\mathcal{R}$, $\mathcal{R}^\mu$, and $\mathcal{R}^{\mu\nu}$ consist of terms of orders $\order{\IRe_{i}\, \IRe_{j}}$, with $i, j \in \{ \Pi, \vec{V}, \tens{\pi} \}$ respectively.
Explicitly, one has
\begin{subequations}\label{eqs:J}
\begin{align}
        \label{transport:J:rel}
    \mathcal{J}
    \coloneqq{}& - \ell_{\Pi V} \nabla_{\mu} V^{\mu}
        - \tau_{\Pi V} V^{\mu} F_{\mu}
        - \delta_{\Pi \Pi} \Pi \theta
        - \lambda_{\Pi V} V^{\mu} I_{\mu}
        + \lambda_{\Pi \pi} \pi^{\mu \nu} \sigma_{\mu \nu} \;, \\
        \label{transport:Jmu:rel}
    \begin{split}
    \mathcal{J}^{\mu}
    \coloneqq{}& - \tau_{V} V_{\nu} \omega^{\nu \mu}
        - \delta_{V V} V^{\mu} \theta
        - \ell_{V \Pi} \nabla^{\mu} \Pi
        + \ell_{V \pi} \Delta^{\mu \nu} \nabla_{\lambda} \pi^{\lambda}_{\nu}
        + \tau_{V \Pi} \Pi F^{\mu}- \tau_{V \pi} \pi^{\mu \nu} F_{\nu}  \\
       & - \lambda_{V V} V_{\nu} \sigma^{\mu \nu}
        + \lambda_{V \Pi} \Pi I^{\mu}
        - \lambda_{V \pi} \pi^{\mu \nu} I_{\nu} \;,
    \end{split} \\
        \label{transport:Jmunu:rel}
    \begin{split}
    \mathcal{J}^{\mu \nu}
    \coloneqq{}& 2 \tau_\pi \pi^{\langle \mu}_{\lambda} \omega^{\nu \rangle \lambda}
        - \delta_{\pi \pi} \pi^{\mu \nu} \theta
        - \tau_{\pi \pi} \pi^{\lambda \langle \mu} \sigma^{\nu \rangle}_{\lambda}
        + \lambda_{\pi \Pi} \Pi \sigma^{\mu \nu}
        - \tau_{\pi V} V^{\langle \mu} F^{\nu \rangle} \\
        & + \ell_{\pi V} \nabla^{\langle \mu} V^{\nu \rangle}
        + \lambda_{\pi V} V^{\langle \mu} I^{\nu \rangle} \;,
    \end{split}
\end{align}
\end{subequations}
as well as
\begin{subequations} \label{eqs:R}
    \begin{align}
            \label{transport:R:rel}
        \mathcal{R}
        &\coloneqq \varphi_1 \Pi^2 + \varphi_2 V^\mu V_\mu + \varphi_3 \pi^{\mu\nu} \pi_{\mu\nu} \;, \\
            \label{transport:Rmu:rel}
        \mathcal{R}^\mu
        &\coloneqq \varphi_4 \pi^{\mu\nu}V_\nu +\varphi_5 \Pi V^\mu \;, \\
            \label{transport:Rmunu:rel}
        \mathcal{R}^{\mu\nu} &\coloneqq \varphi_6 \Pi \pi^{\mu\nu} + \varphi_7 \pi^{\lambda\langle\mu}\pi^{\nu\rangle}{}_\lambda +\varphi_8 V^{\langle\mu} V^{\nu\rangle} \;.
    \end{align}
\end{subequations}

\subsection{Transport properties of the fluid}\label{Transport-properties-of-the-fluid}

\begin{figure}
\includegraphics[scale=1]{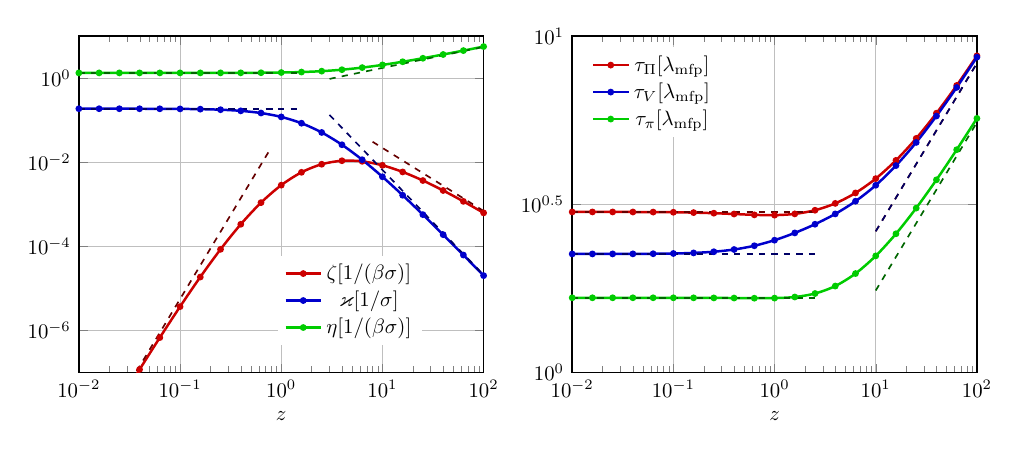}
\caption{The first-order coefficients $\zeta$, $\varkappa$, and $\eta$, as well as the relaxation times $\tau_\Pi$, $\tau_V$, and $\tau_\pi$.}
\label{fig:rel_times}
\end{figure}

\begin{figure}
\includegraphics[scale=1]{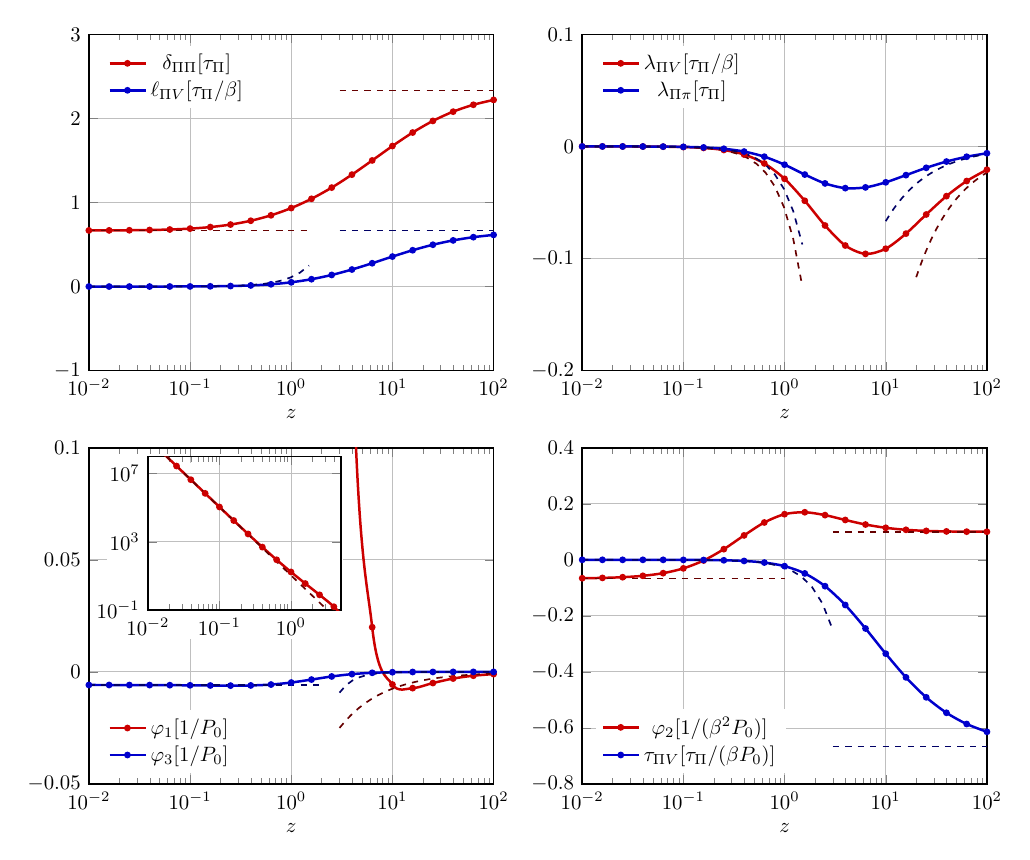}
\caption{The second-order coefficients appearing in the equation of motion for the bulk viscous pressure $\Pi$. The ultra- and non-relativistic limits are indicated by dashed lines.}
\label{fig:coeffs_Bulk}
\end{figure}

\begin{figure}
\includegraphics[scale=1]{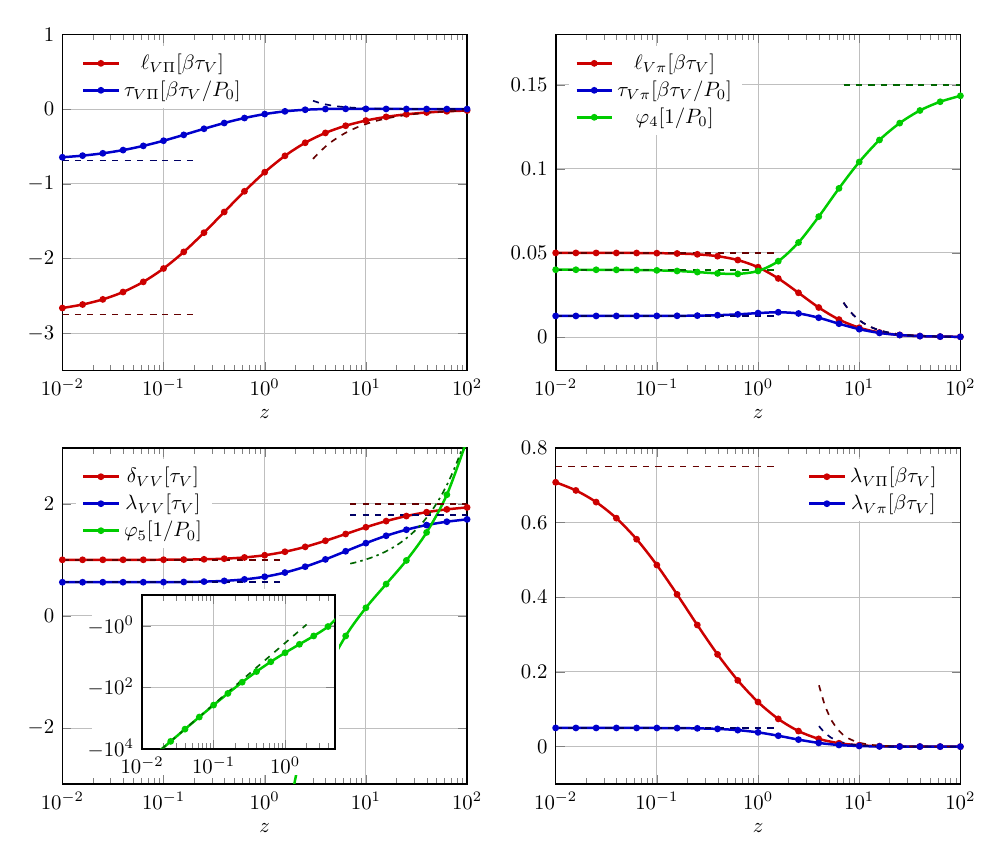}
\caption{The second-order coefficients appearing in the equation of motion for the diffusion current $\vec{V}$.
    The ultra- and non-relativistic limits are indicated by dashed lines.
    Since $\varphi_5$ converges slowly for $z \rightsquigarrow \infty$, we show its next-to-leading order asymptotic, $\varphi_5 \rightsquigarrow z/40+121/160$, as a dash-dotted line.
}
\label{fig:coeffs_n}
\end{figure}

\begin{figure}
\includegraphics[scale=1]{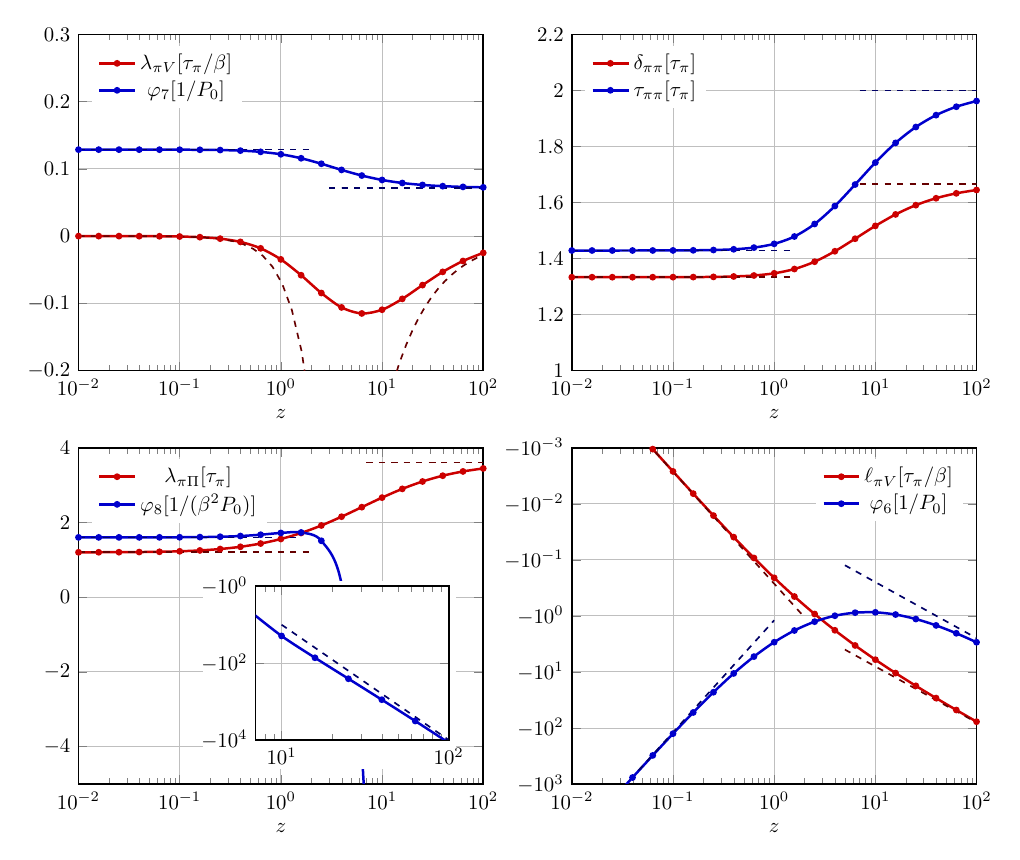}
\caption{The second-order coefficients appearing in the equation of motion for the shear-stress tensor $\tens{\pi}$.
    The ultra- and non-relativistic limits are indicated by dashed lines.
}
\label{fig:coeffs_shear}
\end{figure}

The coefficients appearing in \cref{transport:bulk:rel,transport:diffusion:rel,transport:shear:rel} can be explicitly computed after specifying a collision term.
For a constant cross-section and classical statistics, the results for the bulk viscosity $\zeta$, the thermal conductivity $\varkappa$, the shear viscosity $\eta$, as well as the relaxation times $\tau_\Pi$, $\tau_V$, and $\tau_\pi$ are plotted as functions of $z\coloneqq m\beta$ in \cref{fig:rel_times}.
The second-order coefficients determining the evolution of the bulk viscous pressure, the particle diffusion, and the shear-stress tensor that were introduced in \cref{eqs:J,eqs:R} are shown in \cref{fig:coeffs_Bulk,fig:coeffs_n,fig:coeffs_shear}, respectively.
Therein the ultra- and non-relativistic limits are indicated by dashed lines.
Additionally, the ultra- and non-relativistic limits of all coefficients are given in \cref{tab:coeff_Pi,tab:coeff_n,tab:coeff_pi}.
Note that, since we employ the so-called 14-moment approximation, the coefficients which appear in the terms $\mathcal{J}$, $\mathcal{J}^\mu$, and $\mathcal{J}^{\mu\nu}$ coincide with the results shown in Ref.~\cite{Ambrus:2023qcl}, where the collision term was taken in relaxation-time approximation. On the other hand, the coefficients appearing in the terms $\mathcal{R}$, $\mathcal{R}^\mu$, and $\mathcal{R}^{\mu\nu}$, vanish in the case of the relaxation-time approximation. So far, only $\varphi_4$, $\varphi_7$, and $\varphi_8$ have been computed, and only in the ultra-relativistic limit for a hard-sphere gas, cf. Ref.~\cite{Molnar:2013lta}. Here, we fill this gap by showing all coefficients $\varphi_1$--$\varphi_8$ for a wide range of $z$.
The computation of the coefficients has been carried out in a \textsc{Mathematica} notebook \cite{Mathematica} that can be found, along with the associated data points, in the ancillary files to this article.

\begin{table}
\begin{tabular}{|c|c|c|c|c|c|}
\hline
&$\zeta[1/(\beta\sigma)]$&$\tau_\Pi[\lambda_{\text{mfp}}]$ & $\delta_{\Pi\Pi}[\tau_\Pi]$ & $\ell_{\Pi V}[\tau_\Pi/\beta]$ & $\tau_{\Pi V}[\tau_\Pi/(\beta P_0)]$  \\ \hline
$z \rightsquigarrow 0$& $z^4/18$ & $3$ & $2/3$ & $z^2/9$ & $-z^2/36$ \\\hline
$z \rightsquigarrow \infty$& $25\sqrt{\pi}/(64 z^{3/2})$  & $15\sqrt{\pi}\sqrt{z}/32$ & $7/3$ & $2/3$ & $-2/3$   \\\hline
\end{tabular}
\begin{tabular}{|c|c|c|c|c|c|}
\hline
& $\lambda_{\Pi V}[\tau_\Pi/\beta]$ & $\lambda_{\Pi\pi}[\tau_\Pi]$ & $\varphi_1[1/P_0]$ & $\varphi_2[1/(\beta^2 P_0)]$ & $\varphi_3[1/P_0]$   \\ \hline
$z \rightsquigarrow 0$& $-z^2/18$ & $-7z^2/180$ & $54/(5z^4)$ & $-1/15$ & $-7/1200$\\\hline
$z \rightsquigarrow \infty$& $-7/(3z)$ & $-2/(3z)$ & $-3/(40z)$ & $1/10$ & $-1/(4z^3)$  \\\hline
\end{tabular}
\caption{The first-and second-order coefficients appearing in the equation for the bulk viscous pressure $\Pi$, in the ultra-relativistic ($z \rightsquigarrow 0$) and non-relativistic ($z \rightsquigarrow \infty$) limits.}\label{tab:coeff_Pi}
\end{table}

\begin{table}
\begin{tabular}{|c|c|c|c|c|c|c|c|}
\hline
&$\varkappa[1/\sigma]$&$\tau_V[\lambda_{\text{mfp}}]$ & $\delta_{VV}[\tau_V]$ & $\ell_{V\Pi}[\beta\tau_V]$ & $\ell_{V\pi}[\beta\tau_V]$ & $\tau_{V\Pi}[\beta\tau_V/P_0]$  \\ \hline
$z \rightsquigarrow 0$& $3/16$ & $9/4$ & $1$ & $-11/4$ & $1/20$ & $-11/16$ \\\hline
$z \rightsquigarrow \infty$& $75\sqrt{\pi}/(64 z^{5/2})$ & $15\sqrt{\pi}\sqrt{z}/32$ & $2$ & $-2/z$ & $1/z^2$ & $1/z^2$\\\hline
\end{tabular}
\begin{tabular}{|c|c|c|c|c|c|c|}
\hline
& $\tau_{V\pi}[\beta\tau_V/P_0]$&$\lambda_{VV}[\tau_V]$ & $\lambda_{V\Pi}[\beta\tau_V]$ & $\lambda_{V\pi}[\beta\tau_V]$ & $\varphi_4[1/P_0]$ & $\varphi_5[1/P_0]$ \\ \hline
$z \rightsquigarrow 0$& $1/80$ & $3/5$ & $3/4$ & $1/20$ & $1/25$ & $-18/(5z^2)$\\\hline
$z \rightsquigarrow \infty$ & $1/z^2$& $9/5$ & $21/(2z^3)$ & $7/(2z^3)$ & $3/20$ & $z/40$ \\\hline
\end{tabular}
\caption{The first-and second-order coefficients appearing in the equation for the particle-diffusion current $\vec{V}$, in the ultra-relativistic ($z \rightsquigarrow 0$) and non-relativistic ($z \rightsquigarrow \infty$) limits.}\label{tab:coeff_n}
\end{table}

\begin{table}
\begin{tabular}{|c|c|c|c|c|c|c|c|c|c|c|}
\hline
&$\eta[1/(\beta\sigma)]$&$\tau_\pi[\lambda_{\text{mfp}}]$ & $\delta_{\pi\pi}[\tau_\pi]$ & $\tau_{\pi\pi}[\tau_\pi]$ & $\lambda_{\pi \Pi}[\tau_\pi]$ & $\lambda_{\pi V}[\tau_\pi/\beta]=P_0\tau_{\pi V}$ & $\ell_{\pi V}[\tau_\pi/\beta]$ &$\varphi_6[1/P_0]$ & $\varphi_7[1/P_0]$ & $\varphi_8[1/(\beta^2 P_0)]$ \\ \hline
$z \rightsquigarrow 0$& $4/3$ & $5/3$ & $4/3$ & $10/7$ & $6/5$ & $-z^2/15$ & $-4z^2/15$ & $-6/(5z^2)$ & $9/70$ & $8/5$\\\hline
$z \rightsquigarrow \infty$& $5\sqrt{\pi}\sqrt{z}/16$ & $5\sqrt{\pi}\sqrt{z}/16$ & $5/3$ & $2$ & $18/5$ & $-14/(5z)$ & $-4z/5$ & $-z/40$ & $1/14$ & $-z^3/100$ \\\hline
\end{tabular}
\caption{The first-and second-order coefficients appearing in the equation for the shear-stress tensor $\tens{\pi}$, in the ultra-relativistic ($z \rightsquigarrow 0$) and non-relativistic ($z \rightsquigarrow \infty$) limits. The coefficients $\lambda_{\pi V}$ and $P_0\tau_{\pi V}$ are equal due to the $14$-moment approximation.}\label{tab:coeff_pi}
\end{table}

%% file: part-3.tex
In the context of non-relativistic kinetic theory, the method of moments has also been used to derive fluid-dynamical equations, starting with the seminal work of \citet{Grad:1949zza}.
In this section, we analytically compute the asymptotic regime of relativistic hydrodynamics and compare it to the equations obtained by Grad (in the slightly linearized version derived by \citet{Struchtrup:2004}).

\paragraph{Transport equations for hydrodynamics}

First, in the relaxation \cref{transport:bulk:rel,transport:diffusion:rel,transport:shear:rel}, we exchange the particle-diffusion couplings for those of heat diffusion, and then reduce the resulting equation to the form where a direct comparison with the non-relativistic equations may be made.
Here, as Ref.~\cite[p. 216]{Agrawal:2019} shows, one well known result of hydrodynamics in the framework of kinetic theory -- the 13-moment transport equations for heat diffusion and shear stress due to Grad~\cite{Grad:1949zza} -- reads,
\begin{subequations}
\label{eqs:Grad}
\begin{align}
        \label{eq:Grad:diffusion:pre}
    \begin{split}
    - \frac{2}{3} \frac{P_0}{\eta} q^{i}
    ={}& \dv{q^{i}}{{t}}
        + \frac{5}{2} P_0 \pdv{RT}{{x_{i}}}
        + \frac{7}{5} q^{i} \pdv{u_{k}}{{x_{k}}}
        + q_{k} \pdv{u^{i}}{{x_{k}}}
        + \frac{2}{5} q_{k} \left( \pdv{u^{i}}{{x_{k}}}
            + \pdv{u^{k}}{{x_{i}}} \right) \\
        &- RT \pi^{ik} \pdv{\ln P_0}{{x^{k}}}
        + RT \pdv{\pi^{ik}}{{x^{k}}}
        + \frac{7}{2} \pi^{ik} \pdv{RT}{{x^{k}}}
        - \frac{\pi^{ik}}{\rho_0} \pdv{\pi_{kl}}{{x_{l}}} \;,
    \end{split} \\
        \label{eq:Grad:shear}
    - \frac{P_0}{\eta} \pi^{ij}
    ={}& \dv{\pi^{ij}}{t}
        + 2 P_0 \pdv{u^{\langle i}}{{x_{j \rangle}}}
        + \pi^{ij} \pdv{u_{k}}{x_{k}}
        + 2 \pi^{r \langle i} \pdv{u^{j \rangle}}{{x^{r}}}
        + \frac{4}{5} \pdv{q^{\langle i}}{{x_{j \rangle}}} \;,
\end{align}
\end{subequations}
where $\vec{q}$ is the heat flow, $\rho_0$ the mass density, and $R \coloneqq k_{B}/m \coloneqq 1/m$ the gas constant per unit mass.
Furthermore, $\dd / \dd{t}$ is the (non-relativistic) comoving derivative.
Note that the diffusion \cref{eq:Grad:diffusion:pre} slightly differs from the one given in the reference, but is algebraically equivalent, as shown in \cref{13-moment-diffusion-equation}.

\paragraph{Asymptotic transport equations}

The first step in the process of matching Grad's equations lies in algebraically manipulating \cref{transport:bulk:rel,transport:diffusion:rel,transport:shear:rel} to match the form of \cref{eq:Grad:diffusion:pre,eq:Grad:shear}, which will be shown in \cref{Transport-equations-for-bulk-viscous-pressure-heat-diffusion-shear-stress-tensor}.
Then, in \cref{Asymptotic-thermodynamics}, we discuss some asymptotic relations of thermodynamics.
Finally, in \cref{Matching-the-Grad-like-equations-in-the-NR-limit-to-Grad-s-equations}, inserting the asymptotic values of transport coefficients from \cref{tab:coeff_Pi,tab:coeff_n,tab:coeff_pi} and keeping only the leading-order terms in the mass-over-temperature parameter $z$, the relativistic transport \cref{eq:rel:diffusion:final,eq:rel:shear:final} are shown to be in agreement with the non-relativistic transport \cref{eq:Grad:diffusion:pre,eq:Grad:shear}, if the IReD power-counting scheme is employed consequently.

\subsection{Transport equations for bulk-viscous pressure, heat diffusion, shear-stress tensor}\label{Transport-equations-for-bulk-viscous-pressure-heat-diffusion-shear-stress-tensor}

After substituting $\vec{F} = P_0 \vb*{\nabla} \ln P_0$ into \cref{transport:J:rel,transport:Jmu:rel,transport:Jmunu:rel}, we find
\begin{subequations}
\begin{align}
    \mathcal{J}
    ={}& - \delta_{\Pi \Pi} \Pi \theta
        + \lambda_{\Pi \pi} \pi^{\mu \nu} \sigma_{\mu \nu}
        - \ell_{\Pi V} \nabla_{\mu} V^{\mu}
        - \tau_{\Pi V} P_0 V^{\mu} \nabla_{\mu} \ln P_0
        - \lambda_{\Pi V} V^{\mu} I_{\mu} \:, \\
    \begin{split}
    \mathcal{J}^{\mu}
    ={}& - \tau_{V} V_{\nu} \omega^{\nu \mu}
        - \delta_{V V} V^{\mu} \theta
        - \ell_{V \Pi} \nabla^{\mu} \Pi
        + \ell_{V \pi} \Delta^{\mu \nu} \nabla_{\lambda} \pi^{\lambda}_{\nu}
        + \tau_{V \Pi} \Pi P_0 \nabla^{\mu} \ln P_0
        - \tau_{V \pi} \pi^{\mu \nu} P_0 \nabla_{\nu} \ln P_0 \\
        & - \lambda_{V V} V_{\nu} \sigma^{\mu \nu}
        + \lambda_{V \Pi} \Pi I^{\mu}
        - \lambda_{V \pi} \pi^{\mu \nu} I_{\nu} \;,
    \end{split} \\
    \mathcal{J}^{\mu \nu}
    ={}& 2 \tau_{\pi} \pi^{\lambda \langle \mu} \omega^{\nu \rangle}{}_{\lambda}
        - \delta_{\pi \pi} \pi^{\mu \nu} \theta
        - \tau_{\pi \pi} \pi^{\lambda \langle \mu} \sigma^{\nu \rangle}_{\lambda}
        + \lambda_{\pi \Pi} \Pi \sigma^{\mu \nu}
        - \tau_{\pi V} P_0 V^{\langle \mu} \nabla^{\nu \rangle} \ln P_0
        + \lambda_{\pi V} V^{\langle \mu} I^{\nu \rangle}
        + \ell_{\pi V} \nabla^{\langle \nu} V^{\mu \rangle} \;.
\end{align}
\end{subequations}
\Cref{fluid-dynamics:euler-relation} may be substituted into \cref{fluid-dynamics:duhem-gibbs-relation} to rewrite it in terms of $(\alpha, \beta)$:
\begin{equation}
        \label{beta-from-duhem-gibbs}
    \dd{\beta} = \frac{n_0}{\varepsilon_0 + P_0} \dd{\alpha} - \frac{\beta}{\varepsilon_0 + P_0} \dd{P_0} \;.
\end{equation}
Since $n_0 = \beta P_0$, after introducing $J^\mu \coloneqq \nabla^\mu \beta$, \cref{beta-from-duhem-gibbs} can be written as
\begin{equation}
        \label{pressure-duhem-gibbs}
    \nabla^{\mu} \ln P_0 = I^\mu - h_0 J^\mu \;.
\end{equation}
This equation may be used to represent $\mathbfcal{J} = {\mathbfcal{J}} (\vec{I})$ in terms of ${\mathbfcal{J}} (\vec{J})$.
Furthermore, we introduce the heat flow $\vec{q}$,
\begin{equation} \label{particle-heat-correspondence}
    q^{\mu} \coloneqq W^{\mu} - h_0 V^{\mu} = -h_0 V^\mu \;,
\end{equation}
where the last equality holds in the Landau frame, cf. \cref{fluid-dynamics:Landau-frame}.
\Cref{particle-heat-correspondence} allows to represent $\mathbfcal{J} = {\mathbfcal{J}} (\vec{V})$ instead by $\mathbfcal{J} = {\mathbfcal{J}} (\vec{q})$.
Thus, inserting \cref{pressure-duhem-gibbs}, the equations become
\begin{subequations}
\begin{align}
    \label{transport:J:pre}
    \mathcal{J}
    ={}& - \delta_{\Pi \Pi} \Pi \theta
        + \lambda_{\Pi \pi} \pi^{\mu \nu} \sigma_{\mu \nu}
        + \frac{\ell_{\Pi V}}{h_0} \left( \nabla^{\mu} q_{\mu}
        - \frac{1}{h_0} q^{\mu} \nabla_{\mu} h_0 \right)
        + \frac{\tau_{\Pi V} P_0 + \lambda_{\Pi V}}{h_0} q^{\mu} \nabla_{\mu} \ln P_0
        + \lambda_{\Pi V} q^{\mu} J_{\mu} \;, \\
    \label{transport:Jmu:rel:1}
    \begin{split}
    \mathcal{J}^{\mu}
    ={}& \frac{\tau_{V}}{h_0} q_{\nu} \omega^{\nu \mu}
        + \frac{\delta_{V V}}{h_0} q^{\mu} \theta
        - \ell_{V \Pi} \nabla^{\mu} \Pi
        + \ell_{V \pi} \Delta^{\mu \nu} \nabla_{\lambda} \pi^{\lambda}_{\nu}
        + \left( \tau_{V \Pi} P_0 + \lambda_{V \Pi} \right) \Pi \nabla^{\mu} \ln P_0 \\
        & - \left( \tau_{V \pi} P_0 + \lambda_{V \pi} \right) \pi^{\mu \nu} \nabla_{\nu} \ln P_0
        + \frac{\lambda_{V V}}{h_0} q_{\nu} \sigma^{\mu \nu}
        + \lambda_{V \Pi} \Pi h_0 J^{\mu}
        - \lambda_{V \pi} h_0 \pi^{\mu \nu} J_{\nu} \;,
    \end{split} \\
    \label{transport:Jmunu:rel:1}
    \begin{split}
    \mathcal{J}^{\mu \nu}
    ={}& 2 \tau_{\pi} \pi^{\lambda \langle \mu} \omega^{\nu \rangle}{}_{\lambda}
        - \delta_{\pi \pi} \pi^{\mu \nu} \theta
        - \tau_{\pi \pi} \pi^{\lambda \langle \mu} \sigma^{\nu \rangle}_{\lambda}
        + \lambda_{\pi \Pi} \Pi \sigma^{\mu \nu}
        + \frac{\tau_{\pi V} P_0
            - \lambda_{\pi V}}{h_0} q^{\langle \mu} \nabla^{\nu \rangle} \ln P_0
        - \lambda_{\pi V} q^{\langle \mu} J^{\nu \rangle} \\
        & - \frac{\ell_{\pi V}}{h_0} \left( \nabla^{\langle \mu} q^{\nu \rangle}
        - \frac{1}{h_0} q^{\langle \mu} \nabla^{\nu \rangle} h_0 \right) \;.
    \end{split}
\end{align}
\end{subequations}
The latter two equations may be further simplified by eliminating $\tens{\omega}$ with the help of \cref{Cauchy-Stokes-decomposition}, and substituting the definitions of $\tens{\sigma}$ and $\theta$, such that
\begin{subequations}
\begin{align}
    \label{transport:Jmu:pre}
    \begin{split}
    \mathcal{J}^{\mu}
    ={}& - \frac{\tau_{V}}{h_0} q_{\nu} \nabla^{[ \mu} u^{\nu ]}
        + \frac{\lambda_{V V}}{h_0} q_{\nu} \nabla^{( \mu} u^{\nu )}
        + \frac{\delta_{V V} - \frac{1}{3} \lambda_{V V}}{h_0} q^{\mu} \nabla_{\lambda} u^{\lambda}
        + \left( \tau_{V \Pi} P_0 + \lambda_{V \Pi} \right) \Pi \nabla^{\mu} \ln P_0 \\
        & - \left( \tau_{V \pi} P_0 + \lambda_{V \pi} \right) \pi^{\mu \nu} \nabla_{\nu} \ln P_0
        - \ell_{V \Pi} \nabla^{\mu} \Pi
        + \ell_{V \pi} \Delta^{\mu \nu} \nabla_{\lambda} \pi^{\lambda}_{\nu}
        + \lambda_{V \Pi} h_0 \Pi J^{\mu}
        - \lambda_{V \pi} h_0 \pi^{\mu \nu} J_{\nu} \;,
    \end{split} \\
    \label{transport:Jmunu:pre}
    \begin{split}
    \mathcal{J}^{\mu \nu}
    ={}& \left(\tau_\pi -\frac{\tau_{\pi\pi}}{2}\right)\pi^{\lambda\langle\mu}\nabla^{\nu\rangle}u_\lambda-\left(\tau_\pi +\frac{\tau_{\pi\pi}}{2}\right) \pi^{\lambda \langle \mu} \nabla_{\lambda} u^{\nu \rangle}
        + \left( \frac{\tau_{\pi\pi}}{3} - \delta_{\pi \pi} \right) \pi^{\mu \nu} \nabla_{\lambda} u^{\lambda}
        + \lambda_{\pi \Pi} \Pi \nabla^{\langle \mu} u^{\nu \rangle}
        \\
        & + \frac{\tau_{\pi V} P_0
            - \lambda_{\pi V}}{h_0} q^{\langle \mu} \nabla^{\nu \rangle} \ln P_0  - \lambda_{\pi V} q^{\langle \mu} J^{\nu \rangle}
        - \frac{\ell_{\pi V}}{h_0} \left( \nabla^{\langle \mu} q^{\nu \rangle}
        - \frac{1}{h_0} q^{\langle \mu} \nabla^{\nu \rangle} h_0 \right) \;.
    \end{split}
\end{align}
\end{subequations}
Similarly, the couplings of second order in Reynolds numbers read
\begin{subequations}
\begin{align}
    \label{transport:R:pre}
    \begin{split}
        \mathcal{R}^{\mu}
        ={}& \varphi_1 \Pi^2 + \frac{\varphi_{2}}{h_{0}^2} q^\mu q_\mu + \varphi_3 \pi^{\mu \nu} \pi_{\mu \nu} \;,
    \end{split} \\
    \label{transport:Rmu:pre}
    \begin{split}
        \mathcal{R}^{\mu}
        ={}& - \frac{\varphi_4}{h_0} \pi^{\mu \nu} q_{\nu} - \frac{\varphi_5}{h_{0}} \Pi q^\mu \;,
    \end{split} \\
    \label{transport:Rmunu:pre}
    \begin{split}
        \mathcal{R}^{\mu \nu}
        ={}& \varphi_6 \Pi \pi^{\mu \nu} + \varphi_7 \pi^{\lambda \langle \mu} \pi^{\nu \rangle}{}_{\lambda} + \frac{\phi_8}{h_{0}^2} q^{\langle \mu} q^{\nu \rangle} \;.
    \end{split}
\end{align}
\end{subequations}
\Cref{transport:J:pre,transport:Jmu:pre,transport:Jmunu:pre}, as well as \cref{transport:R:pre,transport:Rmu:pre,transport:Rmunu:pre} will be employed to match the non-relativistic regime, as shown in the following.

From now on, the definitions of $\theta$, $\tens{\sigma}$ and $\vec{J}$ are explicitly inserted.
Furthermore, we employ that for classical particles it holds that $\dd{h_0} = -c_p \beta^{-2} \dd{\beta}$~\cite[p. 70]{Cercignani:2002}, where $c_p = \partial h_0/\partial T \eval_{P_0}$ denotes the specific heat capacity at constant pressure.
Substituting \cref{transport:J:pre,transport:R:rel} into the transient \cref{transport:bulk:rel} of the bulk viscous pressure yields
\begin{subequations}
\begin{align}
        \label{eq:DNMR:bulk:pre}
    \begin{split}
    \dot{\Pi} + \frac{1}{\tau_{\Pi}} \Pi
    ={}& - \frac{\zeta}{\tau_{\Pi}} \nabla_{\lambda} u^{\lambda}
        - \frac{\delta_{\Pi \Pi}}{\tau_{\Pi}} \Pi \nabla_{\lambda} u^{\lambda}
        + \frac{\lambda_{\Pi \pi}}{\tau_{\Pi}} \pi^{\mu \nu} \nabla_{\langle \mu} u_{\nu \rangle}
        + \frac{\ell_{\Pi V}}{\tau_{\Pi}} \frac{1}{h_0} \nabla^{\mu} q_{\mu}+ \frac{\tau_{\Pi V} P_0 + \lambda_{\Pi V}}{\tau_{\Pi}} \frac{1}{h_0} q^{\mu} \nabla_{\mu} \ln P_0 \\
        &
        + \frac{\lambda_{\Pi V}(h_0\beta)^2+c_p\ell_{\Pi V}}{\tau_{\Pi}} \frac{1}{(h_0\beta)^2}q^{\mu} \nabla_{\mu} \beta
        + \frac{\varphi_1}{\tau_\Pi} \Pi^2 + \frac{\varphi_2}{\tau_\Pi h_0^2} q^\mu q_\mu + \frac{\varphi_3}{\tau_\Pi} \pi^{\mu\nu} \pi_{\mu\nu} \;.
    \end{split}
\end{align}
\Cref{transport:diffusion:rel} is written in terms of the heat flow using \cref{particle-heat-correspondence,pressure-duhem-gibbs}, and \cref{transport:Jmu:pre,transport:Rmu:rel} are substituted,
\begin{align}
        \label{eq:DNMR:diffusion:pre}
    \begin{split}
    \dot{q}^{\langle\mu\rangle}
        + \frac{1}{\tau_{V}} q^{\mu}
    ={}& - \frac{\varkappa}{\tau_{V}} h_0 \left( \nabla^{\mu} \ln P_0 + h_0 \nabla^{\mu} \beta \right)
        + q_{\nu} \nabla^{[ \mu} u^{\nu ]}
        - \frac{\lambda_{V V}}{\tau_{V}} q_{\nu} \nabla^{( \mu} u^{\nu )}
        - \frac{\delta_{V V} - \frac{1}{3} \lambda_{V V}}{\tau_{V}} q^{\mu} \nabla_{\lambda} u^{\lambda} \\
        & - \frac{\tau_{V \Pi} P_0 + \lambda_{V \Pi}}{\tau_{V}} h_0 \Pi \nabla^{\mu} \ln P_0
        + \frac{\tau_{V \pi} P_0 + \lambda_{V \pi}}{\tau_{V}} h_0 \pi^{\mu \nu} \nabla_{\nu} \ln P_0
        + \frac{\ell_{V \Pi}}{\tau_{V}} h_0 \nabla^{\mu} \Pi \\
        & - \frac{\ell_{V \pi}}{\tau_{V}} h_0 \Delta^{\mu \nu} \nabla_{\lambda} \pi^{\lambda}_{\nu}
        - \frac{\lambda_{V \Pi}}{\tau_{V}} h_0^2 \Pi \nabla^{\mu} \beta
        + \frac{\lambda_{V \pi}}{\tau_{V}} h_0^2 \pi^{\mu \nu} \nabla_{\nu} \beta
        + \frac{\varphi_4}{\tau_V} \pi^{\mu\nu} q_\nu +\frac{\varphi_5}{\tau_V} \Pi q^\mu \;.
    \end{split}
\end{align}
Finally, substituting \cref{transport:Jmunu:pre,transport:Rmunu:rel} into \cref{transport:shear:rel} yields
\begin{align}
        \label{eq:DNMR:shear:pre}
    \begin{split}
    \dot{\pi}^{\langle\mu\nu\rangle}
        + \frac{1}{\tau_{\pi}} \pi^{\mu \nu}
        ={}& \frac{2 \eta}{\tau_{\pi}} \nabla^{\langle \mu} u^{\nu \rangle}
        +\frac{2\tau_\pi -\tau_{\pi\pi}}{2\tau_\pi}\pi^{\lambda\langle\mu}\nabla^{\nu\rangle}u_\lambda-\frac{2\tau_\pi +\tau_{\pi\pi}}{2\tau_\pi} \pi^{\lambda \langle \mu} \nabla_{\lambda} u^{\nu \rangle}
        +  \frac{\tau_{\pi\pi} - 3\delta_{\pi \pi} }{3\tau_\pi} \pi^{\mu \nu} \nabla_{\lambda} u^{\lambda}\\
        & + \frac{\lambda_{\pi \Pi}}{\tau_{\pi}} \Pi \nabla^{\langle \mu} u^{\nu \rangle}
        + \frac{\tau_{\pi V} P_0
            - \lambda_{\pi V}}{\tau_{\pi}} \frac{1}{h_0} q^{\langle \mu} \nabla^{\nu \rangle} \ln P_0
         - \frac{\lambda_{\pi V}(h_0\beta)^2+c_p\ell_{\pi V}}{\tau_{\pi}} \frac{1}{(h_0\beta)^2}q^{\langle \mu} \nabla^{\nu \rangle} \beta\\
        & - \frac{\ell_{\pi V}}{\tau_{\pi}} \frac{1}{h_0}\nabla^{\langle \mu} q^{\nu \rangle}
        + \frac{\varphi_6}{\tau_\pi} \Pi \pi^{\mu\nu} + \frac{\varphi_7}{\tau_\pi} \pi^{\lambda\langle\mu} \pi^{\nu\rangle}{}_\lambda + \frac{\varphi_8}{\tau_\pi h_0^2} q^{\langle\mu} q^{\nu\rangle} \;.
    \end{split}
\end{align}
\end{subequations}
At this point, we have written the fluid-dynamical equations in a form that can be directly compared to the Grad equations.
First, however, we need to insert the non-relativistic asymptotics of the transport coefficients.

\subsection{Asymptotic regime of relativistic transient hydrodynamics}\label{Asymptotic-regime-of-relativistic-transient-hydrodynamics}

\subsubsection{Asymptotic thermodynamics}\label{Asymptotic-thermodynamics}

Prior to matching the equations in the non-relativistic regime, some approximations must be made.
The thermodynamic integral in \cref{thermodynamic:I} is analyzed in \cref{Deriving-the-asymptotic-expression-for-thermodynamic-integrals}.
In the non-relativistic regime, it may be written as
\begin{equation} \label{thermodynamic:I:asymptotic-regime}
    I_{nq}
    \rightsquigarrow I_{nq}^{(\infty)}\coloneqq  \frac{e^{\alpha-z}}{(2 \pi)^{3/2}\beta^{n+2}}z^{n-q+1/2} \;.
\end{equation}
For the pressure this implies
\begin{equation} \label{eq:P_asymp}
    P_{0}^{(\infty)}
    = \frac{e^{\alpha-z}}{ (2\pi)^{3/2} \beta^4} z^{3 / 2}  \;.
\end{equation}
Furthermore, as shown in Ref.~\cite[eq.~(2.141)]{Rezzolla:2013}, the specific enthalpy can be divided into the rest-mass and non-relativistic contributions,
\begin{equation} \label{enthalphy:separated}
    h_0
    = m c^2
        + m \left( e_0+ \frac{P_0}{\rho_0} \right) \;,
\end{equation}
where $\rho_0 \coloneqq n_0 m$ is the rest-mass density and $e_0 = e_0(v)$ is the kinetic energy density.
    \footnote{The choice of notation for $\varepsilon_0$ and $e_0$ is switched in Ref.~\cite{Rezzolla:2013}.}
Here we temporarily restored the factors of $c$ such that it becomes clear what the dominant contribution is.
In the non-relativistic regime, only the contribution from the rest mass survives, yielding:
\begin{equation} \label{enthalphy:asymptotic-regime}
     h_0 \rightsquigarrow h_{0}^{(\infty)}
     = m c^2\;.
\end{equation}

\subsubsection{Reproducing Grad's equations}\label{Matching-the-Grad-like-equations-in-the-NR-limit-to-Grad-s-equations}

Inserting the transport coefficients listed in \cref{tab:coeff_Pi,tab:coeff_n,tab:coeff_pi}, as well as $h_0^{(\infty)} = z/\beta$, into \cref{eq:DNMR:bulk:pre,eq:DNMR:diffusion:pre,eq:DNMR:shear:pre} yields
\begin{subequations}
\begin{align}
        \label{eq:Pi_NR}
    \begin{split}
    \dot{\Pi} + \frac{1}{\tau_{\Pi}^{(\infty)}} \Pi
    ={}& - \frac{5P_0^{(\infty)}}{6z^2} \nabla_{\lambda} u^{\lambda}
        - \frac{7}{3} \Pi \nabla_{\lambda} u^{\lambda}
        -\frac{2}{3z} \pi^{\mu \nu} \nabla_{\langle \mu} u_{\nu \rangle}
        + \frac{2}{3 z} \nabla^{\mu} q_{\mu}
            - \left(\frac{2}{3z}+ \frac{7}{3z^2} \right) q^{\mu} \nabla_{\mu} \ln P_0^{(\infty)} \\
        & -\left(\frac{7}{3z} -c_p \frac{2}{3z^2}\right) q^{\mu} \nabla_{\mu} \ln\beta
        - \frac{3}{40z\tau_\Pi^{(\infty)}P_0^{(\infty)}} \Pi^2 + \frac{1}{10z^2c^2\tau_\Pi^{(\infty)} P_0^{(\infty)}} q^\mu q_\mu -\frac{1}{4z^3\tau_\Pi^{(\infty)} P_0^{(\infty)}} \pi^{\mu\nu} \pi_{\mu\nu} \;,
    \end{split} \\
        \label{eq:q_NR}
    \begin{split}
    \dot{q}^{\langle\mu\rangle}
        + \frac{1}{\tau_{V}^{(\infty)}} q^{\mu}
    ={}&  - \frac{5}{2} \frac{P_0^{(\infty)}c^2}{z} \left(\nabla^{\mu} \ln\beta+ \frac{1}{z}\nabla^{\mu} \ln P_0^{(\infty)} \right)
        + q_{\nu} \nabla^{[ \mu} u^{\nu ]}
        - \frac{9}{5} q_{\nu} \nabla^{( \mu} u^{\nu )}
        - \frac{7}{5} q^{\mu} \nabla_{\lambda} u^{\lambda} \\
        & - \left(\frac{1}{z} + \frac{21}{2z^2}\right)c^2 \Pi \nabla^{\mu} \ln P_0^{(\infty)}+ \left( \frac{1}{z} + \frac{7}{2 z^2} \right)c^2 \pi^{\mu \nu} \nabla_{\nu} \ln P_0^{(\infty)}
        - 2c^2 \nabla^{\mu} \Pi
        - \frac{c^2}{z} \Delta^{\mu \nu} \nabla_{\lambda} \pi^{\lambda}_{\nu} \\
        & - \frac{21c^2}{2z} \Pi \nabla^{\mu} \ln \beta
        + \frac{7c^2}{2 z} \pi^{\mu \nu} \nabla_{\nu} \ln \beta + \frac{3}{20\tau_V^{(\infty)}  P_0^{(\infty)}} \pi^{\mu\nu}q_\nu + \frac{z}{40 \tau_V^{(\infty)} P_0^{(\infty)}} \Pi q^\mu \;,
    \end{split} \\
        \label{eq:pi_NR}
    \begin{split}
    \dot{\pi}^{\langle\mu\nu\rangle}
        + \frac{1}{\tau_{\pi}^{(\infty)}} \pi^{\mu \nu}
        ={}& 2 P_0^{(\infty)} \nabla^{\langle \mu} u^{\nu \rangle}
        - 2 \pi^{\lambda \langle \mu} \nabla_{\lambda} u^{\nu \rangle}
        - \pi^{\mu \nu} \nabla_{\lambda} u^{\lambda}+ \frac{18}{5} \Pi \nabla^{\langle\mu} u^{\nu\rangle}
        +\frac{14+4c_p}{5 z} q^{\langle \mu} \nabla^{\nu \rangle} \ln\beta \\
        & + \frac{4}{5} \nabla^{\langle \mu} q^{\nu \rangle}  -\frac{z}{40 \tau_\pi^{(\infty)} P_0^{(\infty)}} \Pi \pi^{\mu\nu} + \frac{1}{14\tau_\pi^{(\infty)}  P_0^{(\infty)}}\pi^{\lambda\langle\mu} \pi^{\nu\rangle}{}_\lambda - \frac{z}{100c^2\tau_\pi^{(\infty)} P_0^{(\infty)}} q^{\langle\mu} q^{\nu\rangle} \;.
    \end{split}
\end{align}
\end{subequations}
where we again restored the factors of $c$ to facilitate taking the non-relativistic limit.
When $z = m \beta c^2 \rightsquigarrow \infty$, it can be seen that the asymptotic value of the bulk viscous pressure vanishes as $1/z^2$.
Thus, while staying accurate to second order in Knudsen and inverse Reynolds numbers, one may set the bulk viscous pressure to zero, $\Pi \mapsto 0$,
in accordance with the fact that it vanishes for simple non-relativistic gases \cite[Chap.~VIII]{Vincenti-Kruger:1975}.
In the following, we will employ this fact and not consider \cref{eq:Pi_NR} further.
In order to rewrite some coefficients in \cref{eq:q_NR,eq:pi_NR}, we use that Grad's definition of $R = k_{B}/m$ implies $z = c^2/RT$.
Further, we note that the asymptotic values of $\tau_\pi$, $\tau_V$, and $\eta$ listed in \cref{tab:coeff_n,tab:coeff_pi} allow for the relations
\begin{equation}
        \label{relaxation-time-as-viscosity-to-pressure}
    \tau_{\pi}^{(\infty)} = \frac{\eta^{(\infty)}}{P_0^{(\infty)}} \;,
    \qquad \frac{\tau_{V}^{(\infty)}}{\tau_{\pi}^{(\infty)}} = \frac{3}{2} \implies \frac{1}{\tau_{V}^{(\infty)}} = \frac{2}{3} \frac{P_0^{(\infty)}}{\eta^{(\infty)}} \;.
\end{equation}
Taking note of these facts and substituting $\beta = 1 / T$, \cref{eq:q_NR,eq:pi_NR} simplify to
\begin{subequations}
\begin{align}
    \label{eq:rel:diffusion:final}
    \begin{split}
    \Delta^{\mu}_\nu\dv{q^{\nu}}{{\tau}}
        + \frac{2}{3} \frac{P_0^{(\infty)}}{\eta^{(\infty)}} q^{\mu}
    ={}& \frac{5}{2} P_0^{(\infty)} R \nabla^{\mu} T
        - q_{\nu} \nabla^{\nu} u^{\mu}
        - \frac{2}{5} q_{\nu} \left( \nabla^\nu u^\mu + \nabla^\mu u^\nu \right)
        - \frac{7}{5} q^{\mu} \nabla_{\lambda} u^{\lambda}
        + RT \pi^{\mu \nu} \nabla_{\nu} \ln P_0^{(\infty)} \\
        & - RT \Delta^{\mu \nu} \nabla_{\lambda} \pi^{\lambda}_{\nu}
        - \frac{7}{2} R \pi^{\mu \nu} \nabla_\nu T
        + \frac{1}{10\eta^{(\infty)}} \pi^{\mu\nu} q_\nu \;,
    \end{split} \\
    \label{eq:rel:shear:final}
    \begin{split}
    \Delta^{\mu\nu}_{\alpha\beta}\dv{{\pi}^{\alpha\beta}}{\tau}
        + \frac{P_0^{(\infty)}}{\eta^{(\infty)}} \pi^{\mu \nu}
    ={}& 2 P_0^{(\infty)} \nabla^{\langle \mu} u^{\nu \rangle}
        - 2 \pi^{\lambda \langle \mu} \nabla_{\lambda} u^{\nu \rangle}
        - \pi^{\mu \nu} \nabla_{\lambda} u^{\lambda}
        + \frac{4}{5} \nabla^{\langle \mu} q^{\nu \rangle}
        + \frac{1}{14 \eta^{(\infty)}}\pi^{\lambda\langle\mu}\pi^{\nu\rangle}{}_\lambda \\
        & - \frac{1}{100 RT \eta^{(\infty)}} q^{\langle\mu}q^{\nu\rangle} \;.
    \end{split}
\end{align}
\end{subequations}
Finally, recalling that in the non-relativistic limit the projector $\Delta^{\mu\nu}$ simply projects onto the 3-space orthogonal to the time direction, one may replace $\nabla_\mu \mapsto \partial^i$.
Similarly, we may set $\dd /\dd{\tau} \mapsto \dd / \dd{t}$.
Then, comparing the spatial part of \cref{eq:rel:diffusion:final,eq:rel:shear:final} with Grad's equations,
\begin{subequations}
\begin{align}
        \tag{\ref{eq:Grad:diffusion:pre}}
    \begin{split}
    - \frac{2}{3} \frac{P_0}{\eta} q^{i}
    ={}& \dv{q^{i}}{{t}}
        + \frac{5}{2} P_0 \pdv{RT}{{x_{i}}}
        + \frac{7}{5} q^{i} \pdv{u_{k}}{{x_{k}}}
        + q_{k} \pdv{u^{i}}{{x_{k}}}
        + \frac{2}{5} q_{k} \left( \pdv{u^{i}}{{x_{k}}}
            + \pdv{u^{k}}{{x_{i}}} \right) \\
        &- RT \pi^{ik} \pdv{\ln P_0}{{x^{k}}}
        + RT \pdv{\pi^{ik}}{{x^{k}}}
        + \frac{7}{2} \pi^{ik} \pdv{RT}{{x^{k}}}
        - \frac{\pi^{ik}}{\rho_0} \pdv{\pi_{kl}}{{x_{l}}} \;,
    \end{split} \\
        \tag{\ref{eq:Grad:shear}}
    - \frac{P_0}{\eta} \pi^{ij}
    ={}& \dv{\pi^{ij}}{t}
        + 2 P_0 \pdv{u^{\langle i}}{{x_{j \rangle}}}
        + \pi^{ij} \pdv{u_{k}}{x_{k}}
        + 2 \pi^{r \langle i} \pdv{u^{j \rangle}}{{x^{r}}}
        + \frac{4}{5} \pdv{q^{\langle i}}{{x_{j \rangle}}} \;,
\end{align}
\end{subequations}
shows that both pairs of equation almost coincide, with two differences.
First, there is one term in the diffusion equation \eqref{eq:Grad:diffusion:pre},
\[
    (-1) \frac{\pi^{ik}}{\rho_0} \pdv{\pi_{kl}}{{x_{l}}} \;,
\]
which is neglected in our treatment since it is of $\order{\Kn\, \IRe_{\pi}^2}$ in the chosen power-counting scheme.
This result agrees with the findings of \citet{Struchtrup:2004}, where the same power counting (therein called \enquote{order-of-magnitude} approach) has been used.
Second, there are three terms of $\order{\IRe^2}$, one in \cref{eq:rel:diffusion:final} $\sim \pi^{\mu\nu}q_\nu$, and two in \cref{eq:rel:shear:final} $\sim \pi^{\lambda\langle\mu}\pi^{\nu\rangle}{}_\lambda, \, q^{\langle\mu}q^{\nu\rangle}$.
These terms appear due to the chosen interaction of hard-sphere type.
On the other hand, Grad's equations have been derived under the assumption of a system comprised of Maxwell molecules \cite[p.~206]{Agrawal:2019}, in which case these types of terms vanish \cite[p.~133]{struchtrupbook}.
Even though these non-linear terms differ between hard spheres and Maxwell molecules, the form of the ratio of thermal conductivity (and thus also the relaxation time of the particle diffusion) and viscosity, cf. \cref{relaxation-time-as-viscosity-to-pressure}, stays the same regardless of particle type, as also found by \citet[p. 127]{Cercignani:2002} for the Chapman-Enskog expansion.

%% file: conclusions.tex
In \cref{Formulation-of-relativistic-fluid-dynamics} we calculated all second-order transport coefficients for arbitrary particle masses in the 14-moment approximation.
The results derived there can henceforth be used in numerical frameworks, allowing to include the effects of non-zero particle masses on the transport properties of the fluid.
Furthermore, we have shown in \cref{Bridging-relativistic-and-non-relativistic-hydrodynamics} how to derive the non-relativistic second-order transport equations of Grad (in the version obtained by Struchtrup), starting from the relativistic formulation.
The agreement of the theories in the non-relativistic regime provides an important consistency check and shows that (a variant of) the Grad equations correctly capture the non-relativistic regime of the transport equations for hydrodynamics.

In the future, it could be worthwhile to explore truncations of hydrodynamics at $\order{\IRe^2\, \Kn}$ and higher.
Notably, to match the third-order gradients of third-order hydrodynamics to a kinetic framework, the introduction of further dissipative degrees of freedom is necessary~\cite{deBrito:2023tgb}.
It may then be interesting to compare the relation of such a relativistic theory to the non-relativistic higher-order versions of the Grad equations~\cite{Mueller1998}.
On the other hand, it may be worthwhile to go beyond the 14-moment approximation.
In the massless case, it has been found that the exact values of the transport coefficients can differ substantially from their 14-moment counterparts~\cite{Wagner:2024}, and extending the analysis to the massive case is a logical step to take.

It should be kept in mind that the region of applicability of the constructed theory is constrained by the framework of kinetic theory, cf. \cref{Developing-a-transport-model-for-hydrodynamics}.
Firstly, fluid dynamics requires a large collectivity of the system by the definition of $\Kn$, which may not be present for systems with a large microscopic system length scale $\lambda$.
In particular, the chosen framework of hydrodynamics is only applicable for fluids with weak coupling constants, e.g., gases.
Secondly, the chosen framework of kinetic theory is valid for systems with negligible higher-than-two particle correlations by virtue of the BBGKY hierarchy, which is justified when fluctuations are negligible.
Many attempts of constructing hydrodynamics from kinetic theory do not take fluctuations into account, and the inclusion of these effects, which has been and is being studied~\cite{fox1970contributions,Calzetta:1999xh,Miron-Granese:2020mbf,SoaresRocha:2024afv}, would help improve the accuracy of the second-order hydrodynamic theory.

%% file: acknowledgements.tex
DW acknowledges support by the project PRIN2022 Advanced Probes of the Quark Gluon Plasma funded by \enquote{Ministero dell'Universit\`a e della Ricerca}.
DW and SP are supported by the Deutsche Forschungsgemeinschaft (DFG, German Research Foundation) through the Collaborative Research Center CRC-TR 211 \enquote{Strong-interaction matter under extreme conditions} -- project number 315477589 - TRR 211.
SP thanks Annamaria Chiarini, Amin-ul-Islam Chowdhury, Ashutosh Dash, Masoud Shokri for fruitful discussions.
The authors thank Henri Rosenbaum who initiated this work in his Bachelor's thesis.

%% file: appendix.tex
\section{Grad's 13-moment diffusion equation}\label{13-moment-diffusion-equation}

In this appendix we present an algebraic equivalent to Grad's equation for heat diffusion.
\citet{Grad:1949zza} proposed the 13 moment for heat flow \cite[p.~216]{Agrawal:2019},
\begin{equation}
    \label{eq:Grad:diffusion}
    \begin{split}
    \dv{q^{i}}{{t}}
        +{}& \frac{5}{2} P_0 \pdv{RT}{{x_{i}}}
        + \frac{5}{2} \pi^{ik} R \pdv{T}{{x^{k}}}
        - \frac{\pi^{ik}}{\rho_0} RT \pdv{\rho_0}{{x^{k}}}
        - \frac{\pi^{ik}}{\rho_0} \pdv{\pi_{kl}}{{x_{l}}} \\
        &+ RT \pdv{\pi^{ik}}{{x^{k}}}
        + \frac{7}{5} q^{i} \pdv{u_{k}}{{x_{k}}}
        + \frac{7}{5} q_{k} \pdv{u^{i}}{{x_{k}}}
        + \frac{2}{5} q_{j} \pdv{u^{j}}{{x_{i}}}
    = - \frac{2}{3} \frac{P_0}{\eta} q^{i} \;.
    \end{split}
\end{equation}
Firstly, the first-order dissipative terms can be reshuffled by employing the ideal gas law, $\rho_0 RT=P_0$,
\begin{align}
    \label{eq:Grad:diffusion:thermal}
    \begin{split}
    \frac{5}{2} \pi^{ik} R \pdv{T}{{x^{k}}}
        - \frac{RT}{\rho_0} \pi^{ik} \pdv{\rho_0}{{x^{k}}}
    ={}& \pi^{ik} \left( \frac{5}{2} \pdv{RT}{{x^{k}}}
        - \frac{1}{\rho_0} \pdv{\rho_0 RT}{{x^{k}}}
        +  \pdv{RT}{{x^{k}}} \right) \\
    ={}& \pi^{ik} \left( \frac{7}{2} \pdv{RT}{{x^{k}}}
        - RT \pdv{\ln P_0}{{x^k}} \right) \\
    ={}& \frac{7}{2} \pi^{ik} \pdv{RT}{{x^{k}}}
        - RT \pi^{ik} \pdv{\ln P_0}{{x^{k}}} \;.
    \end{split}
\end{align}
Secondly, the terms that couple the heat flow to gradients of the fluid velocity in \cref{eq:Grad:diffusion} may be sorted as
\begin{equation}
    \label{eq:Grad:diffusion:velocity-gradients}
\frac{7}{5} q_{k} \pdv{u^{i}}{{x_{k}}}
    + \frac{2}{5} q_{k} \pdv{u^{k}}{{x_{i}}}
= q_{k} \pdv{u^{i}}{{x_{k}}}
    + \frac{2}{5} q_{k} \left( \pdv{u^{i}}{{x_{k}}} + \pdv{u^{k}}{{x_{i}}} \right) \;.
\end{equation}
Thus, substituting \cref{eq:Grad:diffusion:thermal,eq:Grad:diffusion:velocity-gradients} into \cref{eq:Grad:diffusion} and reordering the terms yields
\begin{equation} 
    \begin{split}
    \dv{q^{i}}{{t}}
        +{}& \frac{5}{2} P_0 \pdv{RT}{{x_{i}}}
        + \frac{7}{5} q^{i} \pdv{u_{k}}{{x_{k}}}
        + q_{k} \pdv{u^{i}}{{x_{k}}}
        + \frac{2}{5} q_{k} \left( \pdv{u^{i}}{{x_{k}}}
            + \pdv{u^{k}}{{x_{i}}} \right) \\
        &- RT \pi^{ik} \pdv{\ln P_0}{{x^{k}}}
        + RT \pdv{\pi^{ik}}{{x^{k}}}
        + \frac{7}{2} \pi^{ik} \pdv{RT}{{x^{k}}}
        - \frac{\pi^{ik}}{\rho_0} \pdv{\pi_{kl}}{{x_{l}}}
    = - \frac{2}{3} \frac{P_0}{\eta} q^{i} \;.
    \end{split}
\end{equation}

\section{Collision terms}\label{app:coll}

In this appendix we define the linear and non-linear contributions for the moments of the collision term introduced in \cref{eq:coll_0,eq:coll_1,eq:coll_2}.
Since the corresponding analysis has been carried out thoroughly in Ref.~\cite{Molnar:2013lta}, we do not present a full derivation and merely present the results.
Employing the classical approximation, such that $\widetilde{f}_{0\vecc{k}}=1$, the linear terms are given by
\begin{equation}
    \begin{split}
    \mathcal{A}^{(\ell)}_{rn}
    \coloneqq \frac{1}{2(2\ell+1)}\int_f E_{\vecc{k}}^{r-1} k^{\langle\mu_1}\cdots k^{\mu_\ell\rangle}
        \bigg({}& \mathcal{H}_{\vecc{k}n}^{(\ell)}k_{\mu_1}\cdots k_{\mu_\ell}
        + \mathcal{H}_{\vecc{k}'n}^{(\ell)}k'_{\mu_1}\cdots k'_{\mu_\ell} \\
        & - \mathcal{H}_{\vecc{k}_1n}^{(\ell)}k_{1,\mu_1}\cdots k_{1,\mu_\ell}
        - \mathcal{H}_{\vecc{k}_2n}^{(\ell)}k_{2,\mu_1}\cdots k_{2,\mu_\ell} \bigg) \;,
    \end{split}
\end{equation}
where we introduced the short-hand notation:
\begin{equation}
    \int_f
    \coloneqq \int \dd{K}\dd{K'}\dd{K_1}\dd{K_2} W_{\vecc{k}\vecc{k}'\to\vecc{k}_1\vecc{k}_2} f_{0\vecc{k}}f_{0\vecc{k}'} \;.
\end{equation}
The non-linear terms are more complicated and best treated case by case.
For tensor-ranks $0\leq \ell \leq 2$, we have
\begin{subequations}
\begin{align}
    N_{r-1} &= \sum_{m=0}^\infty \sum_{n=0}^{N_m} \sum_{n'=0}^{N_m} \mathcal{C}^{0(m,m)}_{rnn'} \rho_n^{\alpha_1\cdots \alpha_m}\rho_{n',\alpha_1\cdots \alpha_m}\;,\\
    N_{r-1}^{\langle\mu\rangle} &= \sum_{m=0}^\infty \sum_{n=0}^{N_m} \sum_{n'=0}^{N_{m+1}} \mathcal{C}^{1(m,m+1)}_{rnn'} \rho_n^{\alpha_1\cdots \alpha_m}\rho_{n',\alpha_1\cdots \alpha_m}^\mu\;,\\
    N_{r-1}^{\langle\mu\nu\rangle} &= \sum_{m=0}^\infty \sum_{n=0}^{N_m} \sum_{n'=0}^{N_{m+2}} \mathcal{C}^{2(m,m+2)}_{rnn'} \rho_n^{\alpha_1\cdots \alpha_m}\rho_{n',\alpha_1\cdots \alpha_m}^{\mu\nu}+ \sum_{m=1}^\infty \sum_{n=0}^{N_{m+1}} \sum_{n'=0}^{N_{m+1}} \mathcal{D}^{2(m,m)}_{rnn'} \rho_n^{\alpha_2\cdots \alpha_m\langle\mu}\rho_{n',\alpha_2\cdots \alpha_m}^{\nu\rangle}\;,
\end{align}
\end{subequations}
with the coefficients
\begin{align}
    \begin{split}
    \mathcal{C}^{\ell(m,m+\ell)}_{rnn'} \coloneqq \frac{1}{2(1+2m+2\ell)} \int_f E_{\vecc{k}}^{r-1} k^{\mu_1} \cdots k^{\mu_\ell}
    \bigg[{}& \mathcal{H}_{\vecc{k}_1 n}^{(m)} \mathcal{H}_{\vecc{k}_2 n'}^{(m+\ell)} k_1^{\nu_1} \cdots k_1^{\nu_m} k_{2, \langle \mu_1} \cdots k_{2, \mu_\ell}k_{2, \nu_1} \cdots k_{2, \nu_m \rangle} \bigg. \\
    &+ \left(1 - \delta_{\ell0}\right) \mathcal{H}_{\vecc{k}_2 n}^{(m)} \mathcal{H}_{\vecc{k}_1 n'}^{(m+\ell)} k_2^{\nu_1} \cdots k_2^{\nu_m} k_{1, \langle\mu_1} \cdots k_{1, \mu_\ell} k_{1, \nu_1} \cdots k_{1, \nu_m \rangle} \\
    &- \mathcal{H}_{\vecc{k} n}^{(m)}  \mathcal{H}_{\vecc{k}' n'}^{(m+\ell)} k^{\nu_1} \cdots k^{\nu_m} k'_{\langle \mu_1} \cdots k'_{\mu_\ell} k'_{\nu_1} \cdots k'_{\nu_m \rangle} \\
    & \bigg. - \left( 1-\delta_{\ell0} \right) \mathcal{H}_{\vecc{k}' n}^{(m)}  \mathcal{H}_{\vecc{k} n'}^{(m+\ell)} k'^{\nu_1} \cdots k'^{\nu_m} k_{\langle\mu_1} \cdots k_{\mu_\ell} k_{\nu_1} \cdots k_{\nu_m \rangle} \bigg] \;,
    \end{split}
\end{align}
as well as
\begin{align}
    \begin{split}
    \mathcal{D}^{2(m,m)}_{rnn'} \coloneqq \frac{1}{2d^{(m)}} \int_f E_{\vecc{k}}^{r-1} k^{\langle\mu}k^{\nu\rangle}
        \bigg({}& \mathcal{H}_{\vecc{k}_1 n}^{(m)} \mathcal{H}_{\vecc{k}_2 n'}^{(m)} k_{1,\langle\mu} k_1^{\alpha_1} \cdots k_1^{\alpha_m\rangle} k_{2,\langle\nu} k_{2,\alpha_1} \cdots k_{2,\alpha_m\rangle} \bigg. \\
        &\bigg. - \mathcal{H}_{\vecc{k} n}^{(m)}\mathcal{H}_{\vecc{k}' n'}^{(m)}k_{\langle\mu}k^{\alpha_1}\cdots k^{\alpha_m\rangle}k'_{\langle\nu}k'_{\alpha_1}\cdots k'_{\alpha_m\rangle} \bigg) \;,
    \end{split}
\end{align}
with $d^{(1)}=5$ and $d^{(2)}=35/12$.
Note that, in the 14-moment approximation, the expressions become considerably simpler, i.e.,
\begin{subequations}
\begin{align}
    N_{r-1} &\overset{14 \text{m.}}{\mapsto}  \mathcal{C}^{0(0,0)}_{r00} (\rho_0)^2 + \mathcal{C}^{0(1,1)}_{r00} \rho_0^{\alpha}\rho_{0,\alpha} + \mathcal{C}^{0(2,2)}_{r00} \rho_0^{\alpha\beta}\rho_{0,\alpha\beta}\;,\\
    N_{r-1}^{\langle\mu\rangle} &\overset{14 \text{m.}}{\mapsto} \mathcal{C}^{1(0,1)}_{r00} \rho_0\rho_0^\mu +\mathcal{C}^{1(1,2)}_{r00} \rho_0^\alpha\rho_{0,\alpha}^\mu \;,\\
    N_{r-1}^{\langle\mu\nu\rangle} &\overset{14 \text{m.}}{\mapsto} \mathcal{C}^{2(0,2)}_{r00} \rho_0\rho_0^{\mu\nu} + \mathcal{D}^{2(1,1)}_{r00} \rho_0^{\langle\mu}\rho_0^{\nu\rangle}+\mathcal{D}^{2(2,2)}_{r00} \rho_0^{\alpha\langle\mu}\rho_{0,\alpha}^{\nu\rangle}\;.
\end{align}
\end{subequations}

\section{Relativistic hydrodynamics}\label{Relativistic-hydrodynamics}

In this appendix, we list all coefficients appearing in the equations of motion \cref{transport:bulk:rel,transport:diffusion:rel,transport:shear:rel}, as well as derive the asymptotic expression for $I_{nq}$.

\subsection{Transport coefficients}

\allowdisplaybreaks

The coefficients appearing in the equation of motion for the bulk viscous pressure \eqref{transport:bulk:rel} read:\\
\begin{subequations}
\begin{minipage}{.5\textwidth}
\begin{align}
    \zeta &= \frac{m^2}{3} \alpha_0^{(0)} \tau_{00}^{(0)} \;,\\
    \tau_\Pi &= \tau_{00}^{(0)}\;,\\
    \ell_{\Pi V} &= - \frac{m^2}{3} \left( \gamma_{1}^{(1)} - \frac{G_{30}}{D_{20}} \right) \tau_{00}^{(0)} \;, \\
    \tau_{\Pi V} &= \frac{m^2}{3 \left( \varepsilon_0 + P_0 \right)} \left( \pdv{\gamma_{1}^{(1)}}{{\ln \beta}} - \frac{G_{30}}{D_{20}} \right) \tau_{00}^{(0)} \;, \\
    \delta_{\Pi \Pi} &= \left[\frac{1}{3} \left( 2 + m^2 \gamma_{2}^{(0)} \right) - \frac{m^2}{3} \frac{G_{20}}{D_{20}} \right] \tau_{00}^{(0)} \;,
    \end{align}
    \end{minipage}\hfill
    \begin{minipage}{.5\textwidth}
    \begin{align}
    \lambda_{\Pi V} &= - \frac{m^2}{3} \left( \pdv{\gamma_{1}^{(1)}}{{\alpha}} + \frac{1}{h_0} \pdv{\gamma_{1}^{(1)}}{{\beta}} \right) \tau_{00}^{(0)} \;, \\
    \lambda_{\Pi \pi} &= \frac{m^2}{3} \left( \gamma_{2}^{(2)} - \frac{G_{20}}{D_{20}} \right) \tau_{00}^{(0)} \;, \\
    \varphi_1 &= \frac{9}{m^4} \mathcal{C}_{000}^{0(0,0)}\tau_{00}^{(0)}\;,\\
    \varphi_2 &= \mathcal{C}_{000}^{0(1,1)}\tau_{00}^{(0)}\;,\\
    \varphi_3 &= \mathcal{C}_{000}^{0(2,2)}\tau_{00}^{(0)}\;.
\end{align}
\strut
\end{minipage}
\end{subequations}

The coefficients determining the evolution of the particle diffusion current \eqref{transport:diffusion:rel} are given by:\\
\begin{subequations}
\begin{minipage}{.5\textwidth}
\begin{align}
    \varkappa &=  \alpha_0^{(1)} \tau_{00}^{(1)} \;,\\
    \tau_V &= \tau_{00}^{(1)}\;,\\
    \delta_{VV} &= \left( 1 + \frac{1}{3} m^2 \gamma_{2}^{(1)} \right) \tau_{00}^{(1)} \;, \\
    \ell_{V \Pi} &= \left( \frac{\beta J_{21}}{\varepsilon_0+P_0} - \gamma_{1}^{(0)} \right) \tau_{00}^{(1)} \;, \\
    \tau_{V \Pi} &= \frac{1}{\varepsilon_0+P_0} \left( \frac{\beta J_{21}}{\varepsilon_0+P_0}  - \pdv{\gamma_{1}^{(0)}}{{\ln\beta}} \right) \tau_{00}^{(1)} \;, \\
    \ell_{V \pi} &= \left( \frac{\beta J_{21}}{\varepsilon_0+P_0}  - \gamma_{1}^{(2)} \right) \tau_{00}^{(1)} \;,
    \end{align}
    \end{minipage}\hfill
    \begin{minipage}{.5\textwidth}
    \begin{align}
    \tau_{V \pi} &= \frac{1}{\varepsilon_0 + P_0} \left( \frac{\beta J_{21}}{\varepsilon_0+P_0}  -  \pdv{\gamma_{1}^{(2)}}{{\ln\beta}} \right) \tau_{00}^{(1)} \;, \\
    \lambda_{VV} &= \frac{1}{5} \left( 3 + 2 m^2 \gamma_{2}^{(1)} \right) \tau_{00}^{(1)} \;, \\
    \lambda_{V \Pi} &= \left( \frac{1}{h_0} \pdv{\gamma_{1}^{(0)}}{{\beta}} + \pdv{\gamma_{1}^{(0)}}{{\alpha}} \right) \tau_{00}^{(1)} \;, \\
    \lambda_{V \pi} &= \left( \frac{1}{h_0} \pdv{\gamma_{1}^{(2)}}{{\beta}} + \pdv{\gamma_{1}^{(2)}}{{\alpha}} \right) \tau_{00}^{(1)} \;,\\
    \varphi_4 &= \mathcal{C}_{000}^{1(1,2)}\tau_{00}^{(1)}\;,\\
    \varphi_5 &= -\frac{3}{m^2}\mathcal{C}_{000}^{1(0,1)}\tau_{00}^{(1)}\;.
\end{align}
\strut
\end{minipage}
\end{subequations}

Lastly, the transport coefficients appearing in the equation of motion for the shear-stress tensor \eqref{transport:shear:rel} are:\\
\begin{subequations}
\begin{minipage}{.5\textwidth}
\begin{align}
    \eta &=  \alpha_0^{(2)} \tau_{00}^{(2)} \;,\\
    \tau_\pi &= \tau_{00}^{(2)}\;,\\
    \delta_{\pi \pi} &= \left( \frac{4}{3} + \frac{1}{3} m^2 \gamma^{(2)}_{2} \right) \tau_{00}^{(2)} \;, \\
    \tau_{\pi \pi} &= \left( \frac{10}{7} + \frac{4}{7} m^2 \gamma^{(2)}_{2} \right) \tau_{00}^{(2)} \;, \\
    \lambda_{\pi \Pi} &= \left( \frac{6}{5} + \frac{2}{5} m^2 \gamma^{(0)}_{2} \right) \tau_{00}^{(2)} \;,
    \end{align}
    \end{minipage}\hfill
    \begin{minipage}{.5\textwidth}
    \begin{align}
    \tau_{\pi V} &= - \frac{2}{5} \frac{m^2}{\varepsilon_0+P_0} \pdv{\gamma^{(1)}_{1}}{{\ln \beta}} \tau_{00}^{(2)} \;, \\
    \ell_{\pi V} &= - \frac{2}{5} m^2\gamma^{(1)}_{1} \tau_{00}^{(2)} \;, \\
    \lambda_{\pi V} &= - \frac{2}{5} m^2 \left( \frac{1}{h_0} \pdv{\gamma^{(1)}_{1}}{{\beta}} + \pdv{\gamma^{(1)}_{1}}{{\alpha}} \right) \tau_{00}^{(2)} \;,\\
    \varphi_6 &= -\frac{3}{m^2}\mathcal{C}_{000}^{2(0,2)}\tau_{00}^{(2)}\;,\\
    \varphi_7 &= \mathcal{D}_{000}^{2(2,2)}\tau_{00}^{(2)}\;,\\
    \varphi_8 &= \mathcal{D}_{000}^{2(1,1)}\tau_{00}^{(2)}\;.
\end{align}
\strut
\end{minipage}
\end{subequations}
Note that, in the case of classical statistics, the coefficients $\gamma_r^{(\ell)}$ become independent of $\alpha$, and the corresponding derivatives vanish.

\subsection{Deriving the asymptotic expression for thermodynamic integrals}\label{Deriving-the-asymptotic-expression-for-thermodynamic-integrals}

The thermodynamic integral $I_{nq}$ is given by
\begin{align*}
    I_{nq}
    &= \frac{(-1)^{q}}{\dfact{(2q+1)}} \int \dd{K} f_{0\vecc{k}} E_{\vecc{k}}^{n-2q} \left( m^2 - E_{\vecc{k}}^2 \right)^{q} \\
   &= \frac{1}{\dfact{(2q+1)}} e^{\alpha} \int \frac{\dd{\Omega} \dd{\left| {\vecc{k}} \right|}}{(2\pi)^3} \exp \left( - \beta \sqrt{\lvert {\vecc{k}} \rvert^2 + m^2} \right) \left( \left| \vecc{k} \right|^2 + m^2 \right)^{\frac{n - 2q - 1}{2}} \lvert \vecc{k} \rvert^{2q + 2} \;.
\end{align*}
Substituting $y \coloneqq \beta \left| \vecc{k} \right|$ and $z \coloneqq \beta m$, it can be written as
\begin{align}
    I_{nq}
    &= \frac{1}{\dfact{(2q+1)}} \frac{e^{\alpha}}{2\pi^2 \beta^{n+2}} \int_{0}^{\infty} \dd{y} y^{2q+2} \left( y^2 + z^2 \right)^{\frac{n-2q-1}{2}} \exp \left( - \sqrt{ y^2 + z^2 } \right) \;.
\end{align}
First, expanding the square root in the exponential for $y \ll z$ yields
\begin{align}
    \begin{split}
        \exp \left( - \sqrt{y^2 + z^2} \right)
        ={}& \exp \left( -z - \frac{y^2}{2z}  \right) \exp \left[-z \sum_{k=2}^\infty \binom{1/2}{k} \left(\frac{y}{z}\right)^{2k} \right] \\
        ={}& \exp \left( -z - \frac{y^2}{2z}  \right) \sum_{j=0}^\infty \frac{1}{\fact{j}} \left(\frac{y}{z}\right)^{2j} B_j\left[ \frac{y^2}{8z},\cdots , (-1)^{j+1}\frac{\fact{(2j)}}{4^j \fact{(j+1)}} \frac{y^2}{2z} \right] \;,
    \end{split}
\end{align}
where we separated the non-relativistic contribution, and $B_j$ denotes the $j$-th complete exponential Bell polynomial.

Second, the powers of $y^2 + z^2$ are expanded by
\begin{equation}
    \left( y^2 + z^2 \right)^{\frac{n-2q-1}{2}}
    = \sum_{\ell=0}^{\infty} \binom{\frac{n-1}{2} - q}{\ell} y^{2\ell} z^{n-1-2q-2\ell} \;,
\end{equation}
with the sum ending at a finite value if $(n-1-2q)/2 \in \mathbb{N}_0$.
Then, inserting these expansions into the integral yields
\begin{equation}
    \begin{split}
    I_{nq}
    = \frac{1}{\dfact{(2q+1)}} &\frac{e^{\alpha-z}}{2\pi^2 \beta^{n+2}}
        \sum_{\ell=0}^{\infty} \binom{\frac{n-1}{2} - q}{\ell} z^{n-1-2q-2\ell} \\
        &\times \int_{0}^{\infty} \dd{y} ( y^2 )^{q+\ell+1} \exp \left( - \frac{y^2}{2z} \right) \sum_{j=0}^\infty \frac{1}{\fact{j}} \left(\frac{y}{z}\right)^{2j} B_j\left[\frac{y^2}{8z},\cdots , \frac{(-1)^{j+1}}{4^j}\frac{\fact{(2j)}}{\fact{(j+1)}} \frac{y^2}{2z}\right] \;.
    \end{split}
\end{equation}
Expressing the complete Bell polynomials as a sum over the partial Bell polynomials,
\begin{equation}
    B_j(x_1,\cdots ,x_j)= \sum_{k=1}^j B_{j,k}(x_1,\cdots , x_{j-k+1})\;,
\end{equation}
and using that (with $\gamma=-y^2/2z$ and $\delta = -1/4$)
\begin{equation}
    B_{j,k}(\gamma \delta x_1,\cdots , \gamma \delta^{j-k+1} x_{j-k+1})=\gamma^{k}\delta^j B_{j,k}(x_1,\cdots , x_{j-k+1})\;,
\end{equation}
we can compute the integral and arrive at
\begin{align}
\label{eq:Inq_nonrel}
\begin{split}
     I_{nq}
    = \frac{e^{\alpha-z}}{(2\pi)^{3/2} \beta^{n+2}}
        &z^{n-q+1/2}\sum_{\ell=0}^{\infty} \binom{\frac{n-1}{2} - q}{\ell}z^{-\ell} \frac{\dfact{(2\ell+2q+1)}}{\dfact{(2q+1)}}\\
        &\quad\times \left\{1+\sum_{j=1}^\infty\frac{z^{-j}}{4^j \fact{j}} \sum_{k=1}^j (-1)^{j+k}\frac{\dfact{(2j+2k+2\ell+2q+1)}}{2^k \dfact{(2\ell+2q+1)}} B_{j,k}\left[1,\cdots, \frac{\fact{(2j-2k+2)}}{\fact{(j-k+2)}}\right]\right\}\;.
\end{split}
\end{align}

Depending on the proximity to the non-relativistic regime, one may choose the order at which the sums in \mbox{\cref{eq:Inq_nonrel}}, which (likely) define an asymptotic series, are to be truncated.
In the strict non-relativistic limit, one can truncate at the lowest order, setting $\ell=j=0$.
Then, one has
\begin{equation}
    I_{nq}^{(\infty)}= \frac{e^{\alpha-z}}{(2\pi)^{3/2} \beta^{n+2}}z^{n-q+1/2} \;,
\end{equation}
and thus, as written in \cref{eq:P_asymp},
\begin{equation} \label{asymtotic:pressure}
    P_0^{(\infty)}
    = \frac{n_0^{(\infty)}}{\beta}
    =  \frac{e^{\alpha-z}}{(2\pi)^{3/2} \beta^{4}} z^{3/2} \;.
\end{equation}